\documentstyle[prb,aps]{revtex}

\begin{document}

\title{Oxygen Ordering Superstructures and Structural Phase Diagram of YBa$_2$Cu$_3$O$_{6+x}$
Studied by Hard X-ray Diffraction}

\author{M. v. Zimmermann\cite{MvZ}}

\address{Hamburger Synchrotronstrahlungslabor HASYLAB at Deutsches
Elektronen-Synchrotron DESY, Notkestr. 85,\\ D-22603 Hamburg, Germany}

\author{T. Frello, N. H.~Andersen, J. Madsen, M. K\"all, O. Schmidt}

\address{Condensed Matter Physics and Chemistry Department, Ris\o\ National Laboratory,
DK-4000 Roskilde, Denmark}

\author{ T. Niem\"oller, J. R. Schneider}

\address{Hamburger Synchrotronstrahlungslabor HASYLAB at Deutsches
Elektronen-Synchrotron DESY, Notkestr. 85,\\ D-22603 Hamburg, Germany}

\author{H. F. Poulsen}

\address{Materials Research Department, Ris\o\ National Laboratory, DK-4000 Roskilde, Denmark}

\author{Th. Wolf}

\address{Forschungszentrum Karlsruhe, ITP,  D-76021 Karlsruhe, Germany}

\author{R. Liang, P. Dosanjh\, W. N. Hardy}

\address{Department of Physics, The University of British Columbia, Vancouver, British
Columbia, V6T 1Z1 Canada}

\date{\today}

\maketitle



\begin{abstract}

High energy x-ray diffraction is used to investigate the bulk oxygen ordering properties
of YBa$_2$Cu$_3$O$_{6+x}$. Superstructures of Cu-O chains aligned along the $b$ axis and
ordered with periodicity $ma$, along the $a$ axis have been observed. For $x < 0.62$ the
only observed superstructure is ortho-II with $m=2$. At room temperature we find
ortho-III $(m=3)$ for $0.72\le x\le 0.82$, ortho-V $(m=5)$ in a mixed state with ortho-II
at $x=0.62$, and ortho-VIII $(m=8)$ at $x=0.67$. Ortho-II is a 3D ordered structural
phase, the remaining ones are essentially 2D. None of the superstructures develops long
range ordering. Studies of the ortho-II ordering properties in samples prepared with
$x=0.5$ but by different methods show that finite size domains with internal
thermodynamic equilibrium are formed. The crystal perfection as well as the thermal
annealing history determine the domain size. The temperature dependence of the observed
superstructure ordering is investigated explicitly and a structural phase diagram is
presented.

\end{abstract}


\twocolumn

\section{Introduction}

\label{intro} It is now well-established that YBa$_2$Cu$_3$O$_{6+x}$ (YBCO) is
antiferromagnetic (AF) in the tetragonal structural phase for $0 < x < 0.35$, and becomes
superconducting at low temperatures in the weakly distorted orthorhombic phase for $0.35
\leq x < 1$. By high temperature thermal treatment in a suitable oxygen pressure it is
possible to vary the oxygen composition in a continuous way \cite{And90,Sch91}, that
changes the electronic properties from an AF insulator via an underdoped to an optimally
doped high-$T_c$ superconductor with $T_c =93$ K, and a slightly overdoped material for
$x > 0.93$. YBCO has therefore become a major model system for basic studies of
high-$T_c$ superconductivity and it is a leading candidate for technological
applications.

Structural refinement of neutron diffraction data have shown that the variable amount of
oxygen resides in the CuO$_x$ basal plane \cite{Jorg87}. A strong tendency towards
formation and alignment of Cu-O chains gives rise to the orthorhombic distortion below
the temperature dependent transition line between the tetragonal and the orthorhombic
phase. The importance of oxygen ordering for the superconducting properties has been
verified directly from experimental studies where crystals are quenched from the
tetragonal disordered into the orthorhombic ordered phase. Here it is found that $T_c$ of
quenched YBCO is reduced compared to the equilibrium value and increases with time when
the sample is annealed  at room temperature \cite{Cla90,Vea90,Lib93,Mad97}. Equally, it
has been observed that the oxygen ordering of quenched YBCO crystals increases with time
\cite{Sch95}. The consensus is therefore that the Cu-O chain ordering in the CuO$_x$
basal plane controls the charge transfer leading to superconductivity in the CuO$_2$
planes. However, in spite of the very large number of experimental and theoretical model
studies there is still no finite microscopic understanding of how the charge transfer is
controlled by the Cu-O chain length and superstructure ordering. Thus, it is not settled
how the electronic states formed in the Cu-O chains hybridize with the electronic
structure in the CuO$_2$ planes, give rise to the charge transfer and contribute to the
anisotropy of the superconducting properties. A likely explanation is that the available
structural information is not sufficiently detailed and unambiguous because the oxygen
diffusion kinetics is too slow at the temperatures where the superstructures become
stable. It has therefore not been established which superstructures are actually formed
as function of oxygen composition $x$, impurity level and thermal treatment, and how they
influence the electronic states.

The orthorhombic phase found below the tetragonal to orthorhombic phase transition has
the basic ortho-I structure, and it is formed by Cu-O chains that align along the $b$
axis with oxygen on the so-called O(1) site, whereas the sites on the $a$ axis (called
O(5)) are essentially empty. Ortho-I is a 3D long range ordered structure, but in
commonly prepared crystals true long range order is prevented by the formation of
twin-domains with domain size ranging from a few hundred {\AA}ngstr{\o}m to macroscopic
size. Clearly, there is disorder in the ortho-I chain structure for compositions $x <
1.0$. Therefore, in thermodynamic equilibrium ordered superstructures must be formed for
$T \rightarrow 0$. At $x=0.5$ an ideal 3D ortho-II superstructure may in principle be
formed inside the ortho-I twin domains with perfect Cu-O chains on every second $b$ axis
and the remaining ones are empty, {\it i.e.} they contain only Cu. 3D ordering with the
Cu-O chains stacked on top of one another along the $c$ axis has been observed but only
with finite size ordering in all three crystallographic directions.

Electron microscopy techniques have had a leading role in establishing the
superstructures of YBCO \cite{Zan87,Rey89,Wer88,Bey89}. However, the need for
confirmation by bulk structural techniques is generally recognized, because electron beam
heating of thin crystals may change the mobile oxygen content $x$ and generate transient
non-equilibrium surface structures. Also, it is difficult to obtain quantitative details
about the finite size ordering properties and their temperature dependence by these
techniques. The observed superstructures include the Cu-O chain type of ordering as
ortho-II as well as more complex ordering sequences of essentially full (Cu-O) and empty
(Cu) chains with periodicity $m a$ along the $a$ axis and corresponding superstructure
reflections at modulation scattering vectors: $\vec{Q} = (nh_m \; 0 \; 0)$, where
$h_m=1/m$, and $n < m$ are integers, and the coordinates refer to the reciprocal lattice
vectors. Superstructure reflections with $m=2,3,4,5$ and $8$, which we shall call
ortho-II, ortho-III, ortho-IV, ortho-V and ortho-VIII, respectively, have been observed
experimentally. Ideally, these superstructures may be symbolized by their sequence of
full (1) Cu-O and empty (0) Cu chains. Thus, ortho-II is simply (10) with $x=1/2$, and
the ortho-III sequence is (110) with composition $x=2/3$. Ortho-IV has an ordered
sequence of (1110) and composition $x=3/4$, and ortho-V is a sequential ordering of
ortho-II and ortho-III, {\it i.e.} (10110), with composition $x=3/5$. Ortho-VIII is
ortho-V combined with ortho-III to the sequence (10110110) and ideal composition $x=5/8$.
In principle similar structures with full and empty chains interchanged may be stable,
but they have not been observed experimentally.

Superstructures with unit cells $2 \sqrt2 a \times 2 \sqrt2 a \times c$
\cite{Rey89,Ala88,Son91} and $ \sqrt2 a \times 2 \sqrt2 a \times c$ \cite{Zei92}, the
so-called herringbone type, have been reported. However, as we shall discuss in Section
\ref{Discussion} these superstructures are most likely not from oxygen ordering. The
ortho-II and ortho-III superstructures have been verified as bulk structural phases by
x-ray \cite{Sch95,Zei92,Fle88,Pla94,Str98,Pou96,Gry94} and neutron
\cite{Pla94,Bur92,Zei91,Pla95,Sch94} diffraction techniques. The first observation by
x-ray diffraction was made by Fleming {\it et al.} \cite{Fle88} for ortho-II and by
Plakhty {\it et al.} \cite{Pla92} for ortho-III. Similarly, the first neutron diffraction
data were presented by Zeiske {\it et al.} \cite{Zei91} for ortho-II, and by Plakhty {\it
et al.} \cite{Pla95} and Schleger {\it et al.} \cite{Sch95a} for ortho-III. Analysis of
structure factors obtained from a combination of neutron and x-ray diffraction data has
unequivocally shown that the ortho-II and ortho-III superstructures result from oxygen
ordering in Cu-O chains \cite{Zei92,Gry94,Bur92,Zei93,Ho94,Stra98,Sch93,Had94}. However,
relaxation of cations associated with the oxygen chain ordering contributes significantly
to the superstructure intensities. In particular the barium displacement has a strong
influence on the x-ray diffraction intensity. The displacements show only minor variation
with oxygen composition\cite{Pla95,Ho94}. Also, in the ortho-II phase there was found no
significant change in the displacements as function of temperature \cite{Str98}.
Compiling previous room temperature data from x-ray and neutron diffraction studies, we
find that the ortho-II superstructure has been observed for oxygen compositions $0.35 \le
x \le 0.7$ and the ortho-III superstructure for $0.7 \le x \le 0.77$. As we shall present
below, we have observed clear indications of bulk phase ortho-V correlations for $x =
0.62$ and ortho-VIII for $x=0.67$, but we have found no evidence for the ortho-IV
supestructure. Only a few of the previous studies have been carried out above room
temperature \cite{Sch95,Str98,Pou96,Sch95a}.

From previous neutron and hard x-ray diffraction studies it has been inferred that the
finite size ordering of ortho-II results from formation of anti-phase domains inside the
ortho-I twin domains below an ordering temperature of $T_{OII} = 125(5) ^{\circ}$C
\cite{Sch95,Frel97}. Anisotropic superstructure reflections with a Lorentzian-squared
line shape were found in the ordered state \cite{Sch95} as expected from Porod's law:
$S(q) \propto \frac{1}{q^{d+1}}$ for finite size ordered domains with sharp boundaries in
spatial dimension $d$. Studies of the ordering kinetics following a quench in temperature
from the ortho-I into the ortho-II phase have shown a time dependent domain growth that
is algebraic at early times and logarithmic at late times. The characteristic time of the
growth process is activated with an activation energy of 1.4 eV. At 70 $^\circ$C it is
some days and it extrapolates to several years at room temperature. On this basis it was
suggested that the finite size ordering may result from pinning of the domain walls by
impurities or defects, but recent computer simulations have shown that intrinsic slowing
down due to the large effective activation energy for movement of long Cu-O chains may
contribute as well \cite{Mon98}.

In the present paper we report on experimental studies of the oxygen ordering in YBCO
covering the oxygen compositions $0.35 \le x \le 0.87$ and temperatures up to 250
$^\circ$C, by diffraction of high energy synchrotron radiation ($\sim 100$ keV). Thus, our
studies do not include the region at and above the optimal doping level $x \ge 0.93$
where Kaldis {\it et al.} have observed structural anomalies (see {\it e.g.} Ref.
\ref{ref:Kal97}). The high energy x-ray diffraction technique combines the high
penetration power of neutrons with the high momentum space resolution. The penetration
depth of 100 keV x-rays in YBCO is of the order of 1 mm. This assures that we probe the
bulk properties of the samples and are insensitive to oxygen diffusion in and out of the
surface, and studies in sample environments with varying temperatures and controlled
atmospheres are easily accessible. Finally, the synchrotron intensity is so high that
scattering signals down to a factor of 10$^8$ smaller than the fundamental Bragg peaks
can be resolved, and the kinetics of the ordering can be studied with a time resolution
of 1 second. From our studies we present temperature scans of the structure factors of
the superstructures, determine their phase boundaries and the nature of the ordering.
Firstly, we show that the superstructures including ortho-V and ortho-VIII, that Beyers
{\it et al.} \cite{Bey89} have observed by electron microscopy, represent bulk structural
phases. Secondly, we present extensive studies of the ortho-II superstructure ordering in
crystals of different quality and thermal treatment. Finally, we use the present
structural data jointly with data compiled from previous studies to establish a
structural phase diagram that includes the oxygen superstructures and the tetragonal to
orthorhombic (ortho-I) transition temperature, $T_{T-OI}$, obtained by neutron powder
diffraction. We also review and compare with structural findings by other groups and
discuss our results in relation to structural model studies.

The layout of the paper is as follows: In Section~\ref{Experimental} we supply
information about the crystal growth and oxidation of the sample (\ref{Sample
Preparation}), details about the experimental setup (\ref{Instrument}), and the data
analysis (\ref{Analysis}). The experimental results are presented in
Section~\ref{Results}. First we account for the results for the ortho-II superstructure
formation (\ref{ortho-II}). This includes the dependence of the ortho-II correlation
length on crystal quality and thermal treatment of the sample, the phase transition into
the ortho-I phase, and the stability range of the composition $x$. Then we describe the
ordering properties and the stability range of the ortho-III superstructure
(\ref{ortho-III}). In Subsections~\ref{ortho-V} and \ref{ortho-VIII} we present the
properties of the ortho-V and ortho-VIII superstructure ordering, respectively, and a
structural phase diagram of the oxygen ordering is presented in Section~\ref{Phase}. In
Section \ref{Discussion} we discuss our experimental structural results in relation to
other structural findings and their importance for charge transfer, and to theoretical
model descriptions. A concluding summary is given in Section \ref{Summary}.

\section{Experimental details}

\label{Experimental}

\subsection{Sample Preparation}

\label{Sample Preparation}

The single crystals used to study the different superstructure phases and establish the
phase diagram were grown in YSZ (yttria-stabilized zirconia) crucibles by a flux growth
method \cite{Liang92} using chemicals of 99.999 \% purity for Y$_2$O$_3$ and CuO, and
99.997 \% for BaCO$_3$. The impurity level of the crystals has been analyzed by ICP-MS
(Inductively Coupled Plasma Mass Spectroscopy). The Zr content of the crystals was found
to be less than 10 ppm. by weight. The major impurities were Al, Fe and Zn, the sum of
which amounts to less than 0.2 \% atom per unit cell. When optimally doped $(x=0.93)$
these crystals have $T_c = 93.2$ K and the width of the 10 \% - 90 \% diamagnetic
response is $\Delta T_c = 0.3$ K.

Of the four crystals used to study the ortho-II ordering properties as function of sample
purity, crystals \#1, \#3 and \#4 were grown by the flux growth technique described in
Ref. \ref{ref:Wolf89}, and crystal \#2 as described above. Crystal \#1 was used to study
the ortho-II ordering properties when exposed to six different annealing methods.

YSZ crucibles, and chemicals of purity better than 99.99 \% and 99.9 \% were used for
crystals \#1 and \#3, respectively. Crystal \#4 was grown in a SnO$_2$ crucible with
chemicals of purity better than 99.99 \%. The impurity level of crystals \#1, \#3 and \#4
has not been determined directly, but the flux from the crucibles used to grow crystals
\#1 and \#4 has been analyzed. Excessive amounts of Zr and Hf (up to 15000 ppm for Zr and
550 ppm for Hf) were found, but these elements are known to have a very low solubility in
YBCO. Major impurity components were Al  with concentrations 200 ppm and 335 ppm in the
flux from crystal \#1 and crystal \#4, respectively, and similarly for Eu: 115 ppm and
110 ppm. The superconducting transition temperatures and the widths of the transitions
have been determined in the fully oxygenated state $(x=0.99)$. They are: $T_c = 91.0$ K,
$\Delta T_c < 1.0$K for crystal \#1, $T_c = 92.0$ K, $\Delta T_c = 1.5$ K for crystal
\#3, and $T_c = 91.5$ K, $\Delta T_c < 0.5$ K for crystal \#4. The lower $T_c$ and larger
$\Delta T_c$ values of these crystals are not necessarily a consequence of bad crystal
quality but rather a result of overdoping. Thus crystal \#1 has $T_c = 93.5$ K and
$\Delta T_c < 0.5$ K when optimally doped. For all crystals prepared by high purity
chemicals (better than 99.99 \%) it is likely that the impurities come from minority
components of the crucible material or from the furnace walls. All the crystals are
plate-like with thicknesses of $\frac{1}{2}$ to 1 mm, flat dimensions of 1.5 to 3 mm and
weights ranging from 10 to 70 mg.

The oxygen composition of the crystals was changed by use of a gas-volumetric equipment
\cite{And90,Sch91}. For reduction or oxidation the crystals are heated with a suitable
amount ($\approx 10$ g) of YBCO buffer powder in a quartz tube connected to an external
closed volume system. High purity oxygen gas (99.999 \%) is supplied to the system and
the pressure is controlled and monitored by use of high precision absolute pressure
gauges (MKS Baratron) with accuracy and resolution better than 0.01 \%. The crystals and
the powder are wrapped in platinum foil to prevent reaction with the quartz tube. The
closed volume system is made of ultra-high vacuum components and contained in a
thermostatically controlled environment which allows for accurate determination of the
oxygen pressure and oxygen uptake by the powder and the crystals. The water adsorbed in
the system and the materials is removed prior to the preparation by use of a liquid
nitrogen trap and heating the quartz tube to 300 $^\circ$C. At this temperature there is
no reduction of the crystal and the oxygen equilibrium pressure is in any case
sufficiently low that the oxygen does not condense in the trap. The desired oxygen
composition $x$ is usually established by pumping out or adding a suitable amount of
oxygen gas at temperatures between 500 and 600 $^\circ$C. During subsequent cooling the
oxygen pressure is reduced to assure that $x$ stays essentially constant. A final long
time annealing may be performed at lower temperatures to obtain equilibrium between the
powder and the single crystals and develop the oxygen ordering superstructures. For
studies of the structural phase diagram a characteristic procedure to establish the
superstructure is annealing at 80 $^\circ$C for 10 hours and cooling by 1 $^\circ$C/hour
to room temperature where the crystal is stored for more than one week before the
measurements.

If the starting oxygen composition of the buffer powder is known and it is assumed that
the crystals are in equilibrium with the powder, the oxygen composition may be determined
with an accuracy better than $\Delta x =0.02$ by use of the ideal gas law. The resulting
oxygen composition $x$ has been compared with the known values of the oxygen equilibrium
pressure determined by Schleger {\it et al.} \cite{Sch91}, and full agreement has been
established in all cases. Crystals prepared previously by the method have been examined
by neutron diffraction technique and the oxygen composition $x$ determined from
crystallographic analysis of 375 unique reflections were found to be in full agreement
with the values obtained from the gas-volumetry \cite{Cas96}.

\subsection{Instrument}
\label{Instrument}

The experiments were performed on a triple axis diffractometer at the high-energy beam
line BW5 at HASYLAB in Hamburg \cite{Bou98}. The diffractometer operates in horizontal
Laue scattering geometry and is equipped with a Huber 512 Eulerian cradle and a solid
state Ge detector. The insertion device is a high field wiggler with a critical energy of
26.5 keV at the minimum gap of 20 mm. A 1.5 mm copper filter cuts the spectrum below 50
keV, thereby minimizing the heat load on the monochromator crystal. The incident
radiation with an energy in the range of 100 keV has a penetration depth of $\sim 1$ mm
in YBCO samples. For monochromator and analyzer crystals either $(2 \; 0 \; 0)$ SrTiO$_3$
crystals or $(1 \; 1 \; 1)$ Si/TaSi$_2$ crystals \cite{Neu96} were used. Both types of
crystals had a mosaic spread of $\sim 50$'' (arc seconds), resulting in a longitudinal
resolution of 0.0075 \AA$^{-1}$ at the $(2 \; 0 \; 0)$ reflection of
YBa$_2$Cu$_3$O$_{6+x}$. The transverse resolution is limited by the sample mosaicity,
which was in the range of 0.05$^{\circ}$-0.1$^{\circ}$ for our samples, corresponding to
$\sim 0.0015$ \AA$^{-1}$ at the $(2 \; 0 \; 0)$ reflection. The vertical resolution
depends on the setting of the slits before and behind the sample. They were usually set
to integrate the scattering over a quarter of a reciprocal lattice unit, that is 0.40
\AA$^{-1}$ along the $a$ and $b$ axes and 0.13 \AA$^{-1}$ along the $c$ axis. The sample
was wrapped in Al-foil and mounted in a small furnace, designed for use in an Eulerian
cradle. The furnace temperature was stable within 1$^{\circ}$C. An inert atmosphere of
0.3 bar Ar was introduced into the furnace to prevent oxidation of the crystals. From the
gas-volumetric preparations it is established that the reduction is negligible for
temperatures below 300 $^\circ$C, and we observed no changes in the structural properties
that could be related to a change of oxygen composition during temperature cycling at
temperatures below 250$^\circ$C.

\subsection{Analysis of superstructure data}

\label{Analysis}

The ortho-II superstructure reflections are well described by the scattering function
\begin{equation}
\label{sq}
  S({\bf q})=A/(1+(q_h/\Gamma_h)^2+(q_k/\Gamma_k)^2+(q_l/\Gamma_l)^2)^y
\end{equation}
where $q_i$, $i=h,k,l$ is the reduced momentum transfer and $\Gamma_i$ the reduced
inverse correlation length, related to the correlation length $\xi_i$ by $\xi_h =
a/(2\pi\Gamma_h)$ for the $a$ direction and analogous along $b$ and $c$. Equation
\ref{sq} is a 3D anisotropic Lorentzian raised to the power $y$. The scattering function
$S({\bf q})$ has been convoluted with the resolution function, which is treated as a
$\delta$-function in the scattering plane and an integrating function in the
perpendicular direction. The exponent $y$ indicates the distribution of domain size, {\it
i.e.} $y=1$ points to an exponential decrease of the pair correlations as for example in
the ramified clusters typically for critical fluctuations above the transition
temperature. The exponent $y=2$ may result from a domain size distribution around an
average value $\Gamma_i$ \cite{Fii95} and, as mentioned in the Introduction, the
asymptotic behaviour for large $q$ is in agreement with Porod's law for scattering from
3D finite size domains with sharp boundaries. Furthermore Bray has shown that the tail of
the scattering function from a topological defect of dimension $m$, in a system of
dimension $d$ is given by $S(q)\propto \frac{1}{q^{2d-m}}$ \cite{Bra94}. The relation
between $\Gamma$ and the peak width (HWHM=$\Delta$) is $\Delta/\sqrt{2^{1/y}-1}$. When
full integration of the superstructure peak is performed perpendicular to the scattering
plane by relaxing the vertical aperture the in-plane scattering function derived from Eq.
\ref{sq} is described by a Lorentzian to the power $y'=y-\frac{1}{2}$.

The ortho-III, ortho-V and ortho-VIII superstructures are essentially 2D ordered giving
rise to significant overlap of the peaks along $l$ \cite{Pla95,Pla92,Sch95a}. In this
case full integration in the vertical direction cannot be obtained when the $c$ axis is
perpendicular to the scattering plane. For the 2D ordered superstructures with finite
domain size and sharp boundaries it is expected that the scattering function in the $ab$
plane should be a Lorentzian to the power $y'=y=3/2$ because the integration along $c$ is
rather incomplete, whereas it should be a simple Lorentzian ($y'=y-1/2=1$) when the $c$
direction is in the scattering plane and full integration along the $a$ or $b$ direction
is performed.

\section{Results}
\label{Results}

\subsection{The ortho-II superstructure}
\label{ortho-II}

The ortho-II phase is observed as a 3D ordered structure with anisotropic correlation
length at compositions $0.35\le x < 0.62$ and no other superstructure was found in this
range of oxygen content. Especially at $x=0.35$ the q-space was surveyed unsuccessfully
for correlations of the herringbone,  $\sqrt2 a \times 2\sqrt2 a \times c$, or $2\sqrt2 a
\times 2\sqrt2 a \times c$ type structures which were observed in other studies by
electron \cite{Rey89,Ala88,Cha87}, x-ray \cite{Zei92a} and neutron diffraction
\cite{Son91}.

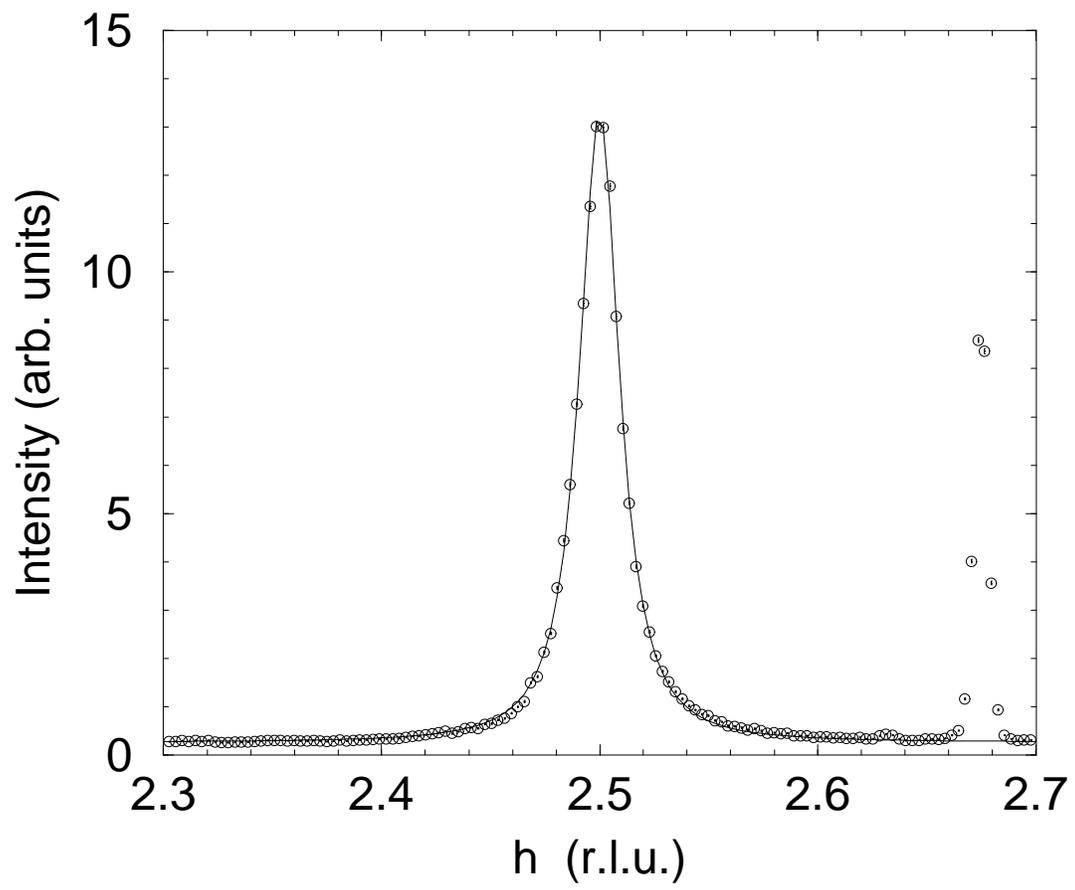
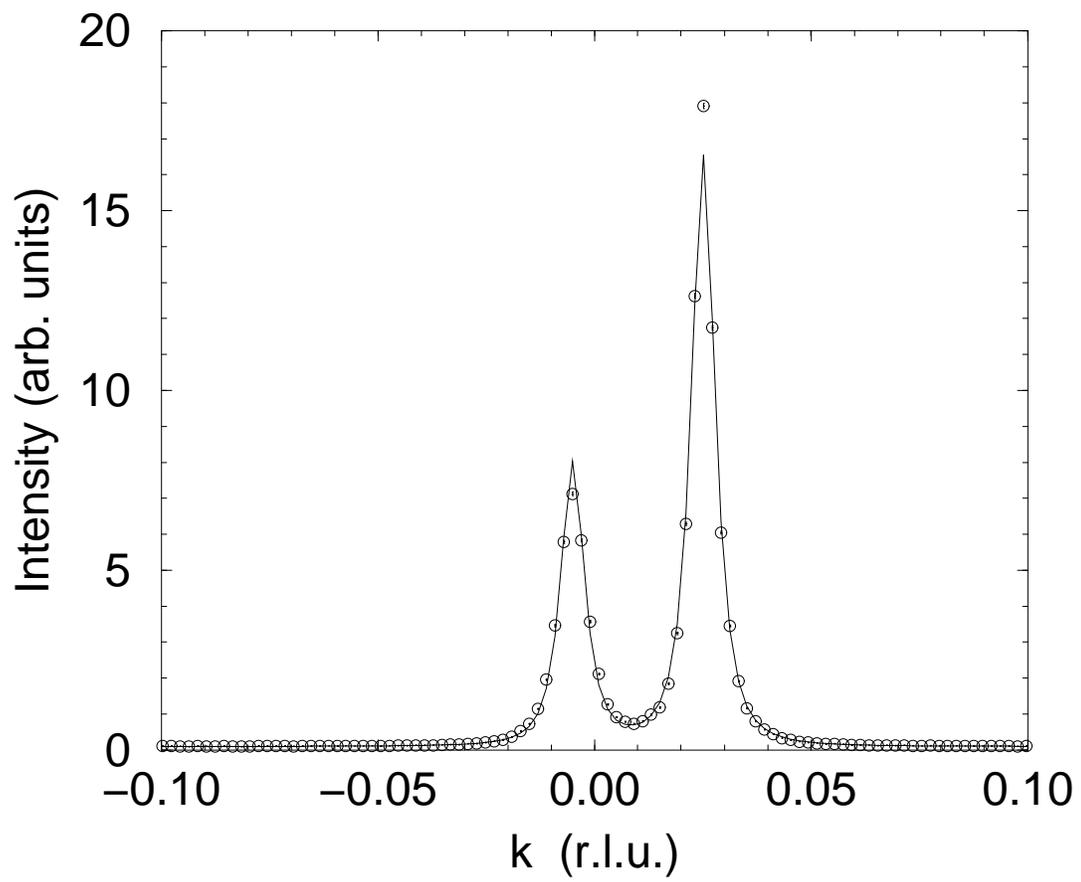
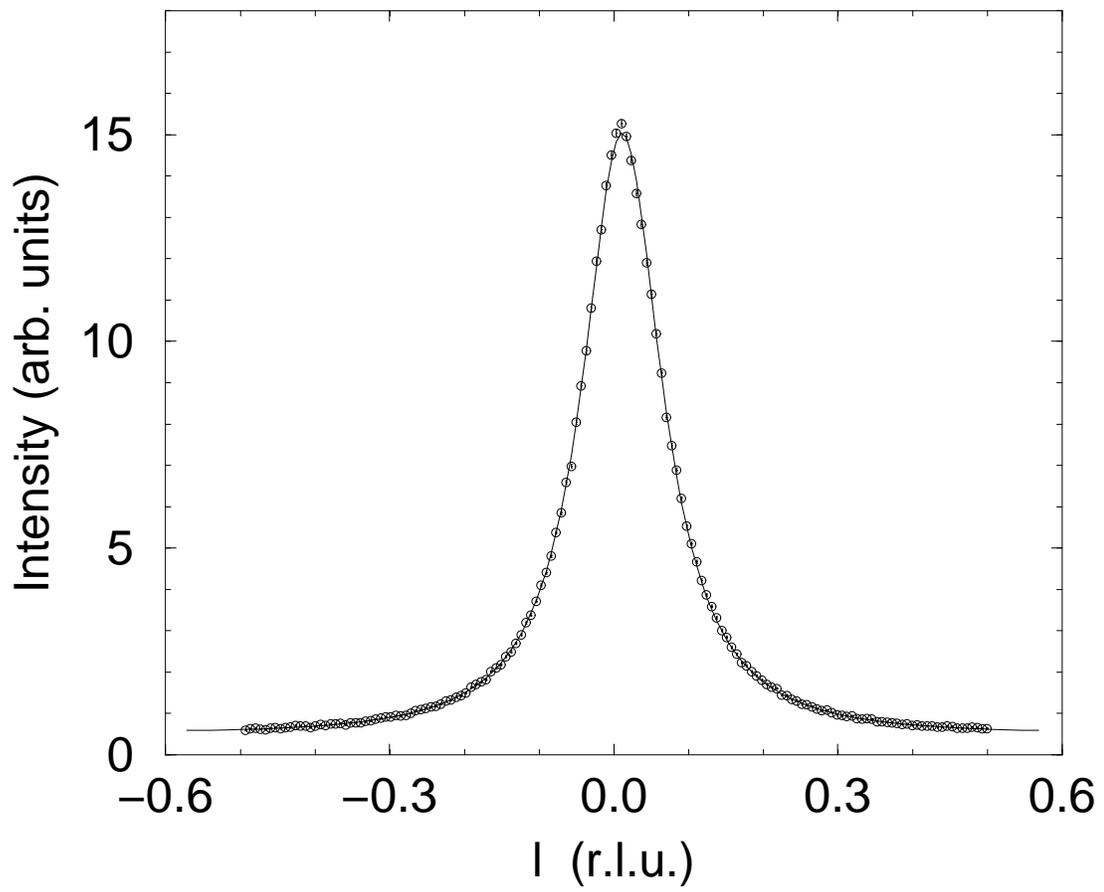
\begin{figure}
  \caption{$(2.5 \; 0 \; 0)$ superstructure reflection of the ortho-II phase at room temperature scanned
  along $h$ (left), $k$ (middle) and $l$ (right) in crystal \#1. The thermal treatment of the crystal corresponds to
  state 2 as described in the text. The lines are least square fits to equation (\ref{sq}). The peak at $h=2.68$ is
  an Al $(2 \; 2 \; 0)$ powder-line.}
  \label{o2-peak}
\end{figure}

The largest correlation lengths are obtained for the ortho-II phase at $x=0.50$. An
example of the $(2.5 \; 0 \; 0)$ superstructure reflection measured in crystal \#1 is
shown in Fig. \ref{o2-peak}. The crystal has been annealed by the thermal procedure
marked 2 below. The room temperature properties of the oxygen superstructures are
strongly dependent on the crystal quality and the thermal treatment. The four crystals
labeled \#1 - \#4 in section \ref{Sample Preparation} have been prepared with $x =0.50$,
annealed at 500 $^\circ$C for 6 days, cooled by 10 $^\circ$C/hour to 100 $^\circ$C where
they were annealed for 36 hours and then quenched to room temperature. The diffraction
studies were performed 5-10 days later. With this thermal treatment we assume, on the
basis of the results presented below and in Ref. \ref{ref:Kal98}, that all the crystals
have reached the late state of ortho-II domain growth, and the influence of the different
room temperature annealing times is considered to be small compared to the differences
due to impurities and defects.

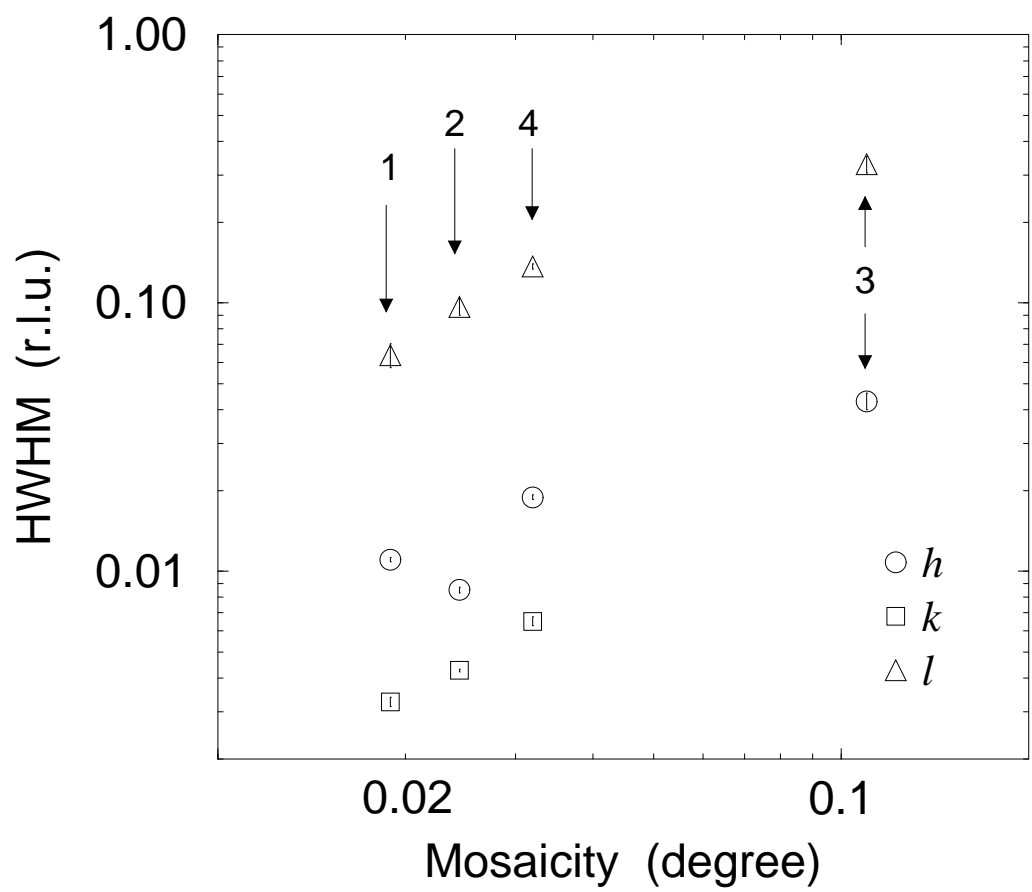
\begin{figure}
  \caption{Half width half maximum (HWHM) of the ortho-II superlattice peak along $h$, $k$ and $l$
plotted as function of sample mosaicity in the $ac$-plane of four differently grown crystals. The growth techniques are described in the text.}
  \label{Defects}
\end{figure}

Although it is possible to determine the impurity level in the crystal, it is often not
known on which lattice sites the various impurities are located and what influence they
have on the oxygen ordering. However, a crystal prepared as described in Ref.
\ref{ref:Wolf89} using Al$_2$O$_3$ as crucible material resulted in a crystal with 6 mole
\% Al, which by neutron diffraction studies was found to be located on the Cu(1) site in
the basal plane \cite{Brecht95}. This crystal did not show ortho-II superstructure
ordering when it was prepared with $x=0.50$. The nature and influence of lattice defects
are equally difficult to quantify. One way to measure the overall quality of the crystal
lattice is the mosaicity, {\it i.e.} the width of the rocking scan of the sample. Fig.
\ref{Defects} shows the HWHM of $h, k$ and $l$ scans obtained at room temperature of the
ortho-II superstructure reflection versus the mosaicity for the four crystals \#1 - \#4.
The $h$ and $l$ scans are measured in the $ac$ scattering plane, and the scan along $k$
in the $ab$ plane. In both cases a full integration of the respective vertical widths,
$\Delta k$ and $\Delta l$ is performed. It is immediately obvious that crystal \#3
prepared with low purity chemicals have the largest mosaicity and the broadest ortho-II
peaks. Thus, the purity of the chemicals is crucial for the development of large ortho-II
domains. For crystals \#1, \#2 and \#4 a linear relation between mosaicity and the HWHM
of the ortho-II superlattice peak is found, with small deviations of the width along $h$.
With the mosaicity as a criteria of crystal quality crystal \#1 is the most perfect one.
This is corroborated by magneto-optic studies of the magnetic flux flow in the crystals.
Crystal \#1 is the only one that shows flux flow instability, which is considered to be a
signature of very high crystal quality \cite{Frello99}.

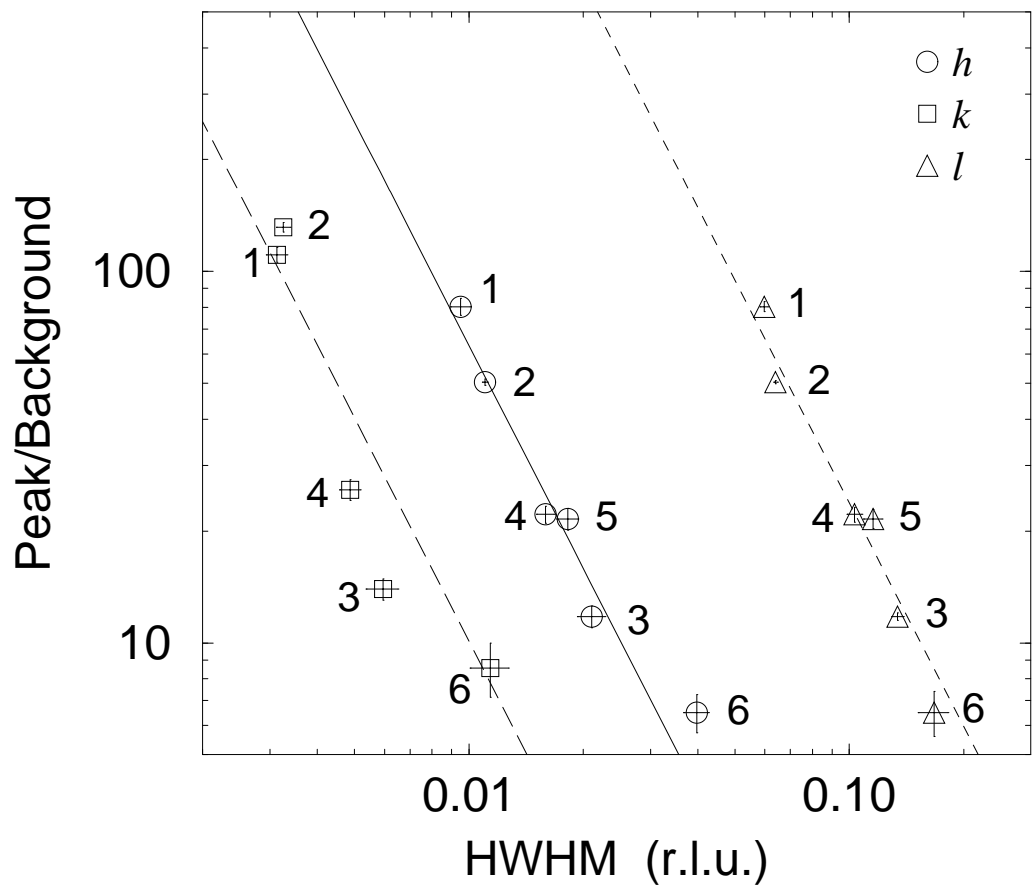
\begin{figure}
  \caption{Peak to background ratio of the ortho-II superlattice reflection $(2.5 \; 0 \; 5)$ plotted as
    function of the peak width measured at crystal \#1 with $x=0.50$ after the different thermal treatments
    (labeled 1 ... 6) that are described in the text. The lines are fits to a quadratic dependence of the
    intensity on the width.}
  \label{peak-HWHM-prep}
\end{figure}

The influence of the thermal treatment has been studied in crystal \#1. After the
preparation for $x=0.50$ mentioned above the crystal has been annealed the following six
ways:
\begin{enumerate}
\item 70 days annealing at 80$^{\circ}$C, cooling to room temperature in steps of 1$^{\circ}$C/hour
\item 5 hours annealing at 100$^{\circ}$C, quenched to room temperature and stored for 10 days
\item quenched from 170$^{\circ}$C to room temperature and stored at room temperature for 97 days
\item cooled down from 170$^{\circ}$C to room temperature in steps of 10$^{\circ}$C every 10 minutes
\item cooled down from 170$^{\circ}$C to room temperature with 4$^{\circ}$C/minutes
\item quenched from 170$^{\circ}$C to room temperature within 3 minutes
\end{enumerate}
In Fig. \ref{peak-HWHM-prep} the normalized peak intensity is plotted versus the inverse
correlation lengths measured at room temperature at the $(2.5 \; 0 \; 5)$ superstructure
reflection for the differently treated samples. The normalization of the peak intensity
is made relative to the background. As mentioned above, the $h$ and $l$ scans are
measured in the $ac$ plane, the scans along $k$ in the $ab$ plane, and a full integration
along the vertical widths is performed. The relation between the measured peak intensity
and the peak widths in all three directions follows a quadratic dependence as indicated
by the lines. This shows that the ratios between the line-widths $\Gamma_h / \Gamma_k$
and $\Gamma_l / \Gamma_k$ are independent of the thermal treatment. Further, since the
measured peak intensity, $I_{peak}^{obs}$, includes an integral over the direction
perpendicular to the scattering plane we establish that also the total integrated
intensity, $I_{int}$, is independent of thermal treatment. This is easily seen from Fig.
\ref{peak-HWHM-prep} and the following relations, where the integration is assumed to be
along $k$:
\begin{eqnarray*}
I_{int} & \propto & I_{peak} \Gamma_h \Gamma_k \Gamma_l\\ I_{peak}^{obs} & \propto &
I_{peak}\Gamma_k \propto I_{int}/(\Gamma_h \Gamma_l) \propto I_{int}/\Gamma_h^2
\end{eqnarray*}
Accordingly, only the correlation length of the superstructure and thereby the
characteristic domain size depend on the sample treatment. This indicates that the finite
size domains have internal thermodynamic order and fill the crystal. Studies of the time
dependent oxygen ordering following a temperature quench from the ortho-I into the
ortho-II phase at this composition confirm that the integrated intensity depends only on
the temperature. These results will be published elsewhere \cite{Kal98}.

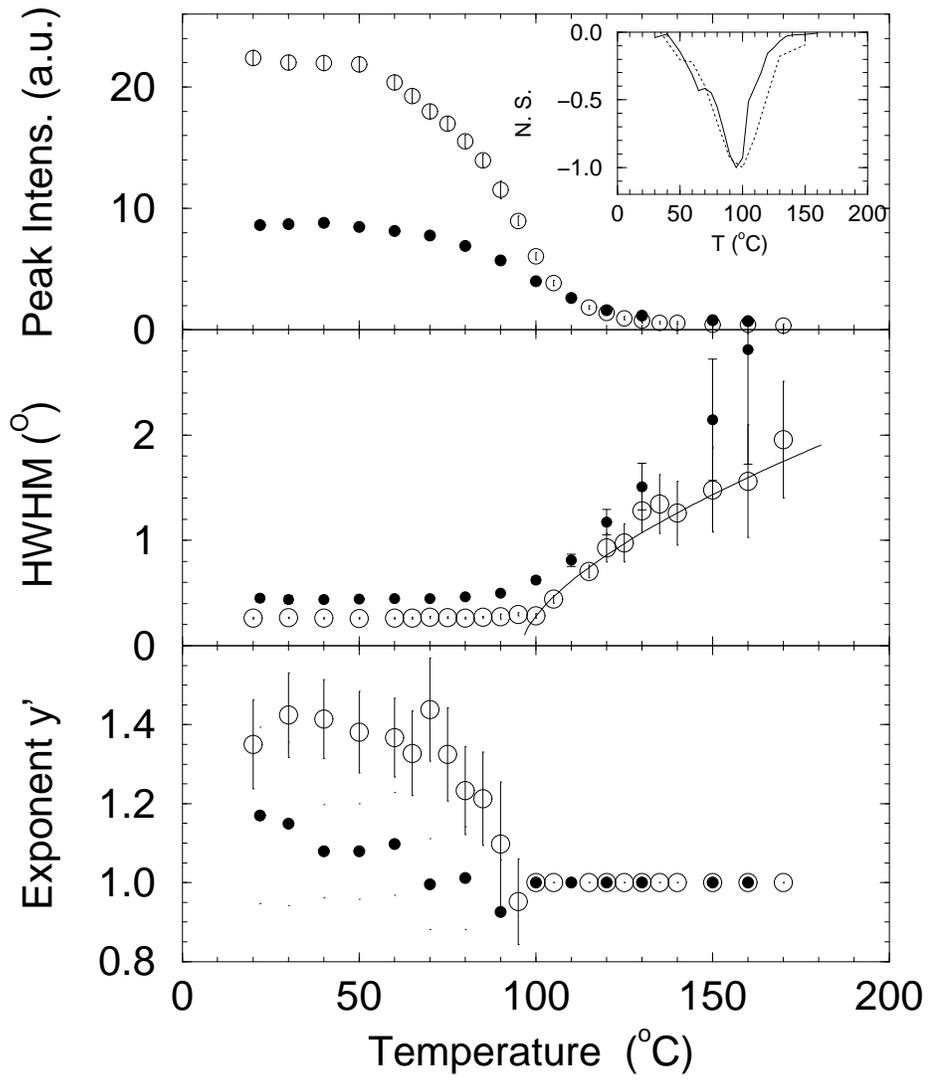
\begin{figure}
  \caption{Top: $(2.5 \; 0 \; 5)$ peak intensity measured by $\omega$-scans for crystal \#1 with
  $x=0.50$ after annealing at 80 $^\circ$C for 70 days and cooling to room temperature by
  1 $^\circ$C/hour (annealing procedure 1 in Subsection \ref{ortho-II}). Open symbols mark the data obtained during
   heating, closed symbols those from cooling. The inset shows the slope of the data for heating (solid line) and
   those for cooling (dashed line)normalized to unity. Middle: Half width half maximum of the superstructure reflection.
   The line is a fit to the critical behavior of a 3-dimensional Ising model. Bottom: Exponent of the
    Lorentzian scattering function. The exponent of the data above 95$^{\circ}$C are fixed to $y'=1$.}
  \label{hihi_trans}
\end{figure}

It is instructive to consider the temperature variation of the ortho-II structure for
crystal \#1 prepared by thermal treatment 1. The phase transition from the ortho-II phase
into the ortho-I phase was studied by means of $\omega$-scans at the $(2.5 \; 0 \; 5)$
reflection during initial heating and at subsequent cooling within one hour. The result
is shown in Fig. \ref{hihi_trans}. Clearly, the peak intensity is lower and the $\omega$
width is larger on cooling than on heating. However, the integrated intensity calculated
as $I_{peak}^{obs} \times \omega^2$ is found to be the same during heating and cooling.
Since the results of the ortho-II superstructure ordering indicate that internal
superstructure order is established inside the finite size domains it is appropriate to
define a transition temperature, $T_{OII}$. Several criteria may be used. Firstly the
variation of the peak intensity of the superstructure reflection plotted in the top part
of Fig. \ref{hihi_trans} shows an inflection point at 95$^{\circ}$C, as determined by the
minimum of the normalized slope (N.S.) of the peak intensity, plotted in the inset. The
inflection point of the peak intensity indicates the cross-over from static order to
critical fluctuations, {\it i.e.} the transition temperature $T_{OII}$. Secondly the
onset of the line broadening of the superstructure reflection also marks the transition
temperature. The temperature dependence of the line width above the transition can be
well described by the critical exponent $\nu=0.63$ for the 3D Ising model:

\begin{eqnarray}
  \label{nu}
  \Delta (T) = \Delta_0^\pm |T-T_c|^{\nu} \quad ,
\end{eqnarray}
where $\Delta_0^\pm$ are the slopes below and above the transition temperature. This
behavior is shown in the middle part of Fig. \ref{hihi_trans}. From the fit to the
heating data in the critical region we find $T_{OII} = 95$ $^\circ$C. Thirdly the line
shape changes from approximately a Lorentzian squared, {\it i.e.} $y'=\frac{3}{2}$, at
room temperature to a simple Lorentzian at the transition temperature: $T_{OII} = 95
^{\circ}$C shown in the bottom part of Fig. \ref{hihi_trans}. A line shape described by a
simple Lorentzian is characteristic for critical fluctuations above the transition
temperature. During cooling a drastic slowing down of the ordering process is observed at
temperatures close to the transition temperature as seen in the variation of the peak
intensity, which starts to deviate from the heating data at $\sim 105^{\circ}$C.

The variation of the peak intensity and the peak width with temperature leads to the
distinction of three areas during the heating cycle. Between room temperature and
50$^{\circ}$C both the peak intensity and the width of the superstructure reflections are
constant, {\it i.e.} both the domain size and the integrated intensity, which is a
measure of the order parameter of the ortho-II phase, are constant within the time period
studied. Between 50$^{\circ}$C and 95$^{\circ}$C the width of the superstructure
reflections is still constant, but the peak intensity decreases with increasing
temperature, indicating that the ortho-II order inside the domains and thereby the number
of oxygen atoms ordered in alternating full and empty chains start to decrease. Finally,
the increasing width and the decreasing intensity in the temperature range above
95$^{\circ}$C indicate the area of critical fluctuations above the transition
temperature. This range is well described by the critical exponents. In contrast, the
behavior below the transition temperature shows substantial deviations from what is
expected from a regular second order phase transition, where a long range ordered phase
should be formed.

The investigation of the temperature dependence of the ortho-II phase at $x=0.42$
exhibits exactly the same behavior as found at $x=0.50$ with a small shift in the
transition temperature. Thus a stoichiometric oxygen content is only of minor importance
for the behavior at the ortho-I/ortho-II phase transition. However, the peak widths,
$\Delta h = 0.031(1), \Delta k = 0.010(1), \Delta l= 0.14(1)$, are significantly larger
than found for the high purity crystal \#1 with $x=0.50$, {\it cf.} Figs. \ref{Defects}
and \ref{peak-HWHM-prep}.

\subsection{The ortho-III structure}
\label{ortho-III}

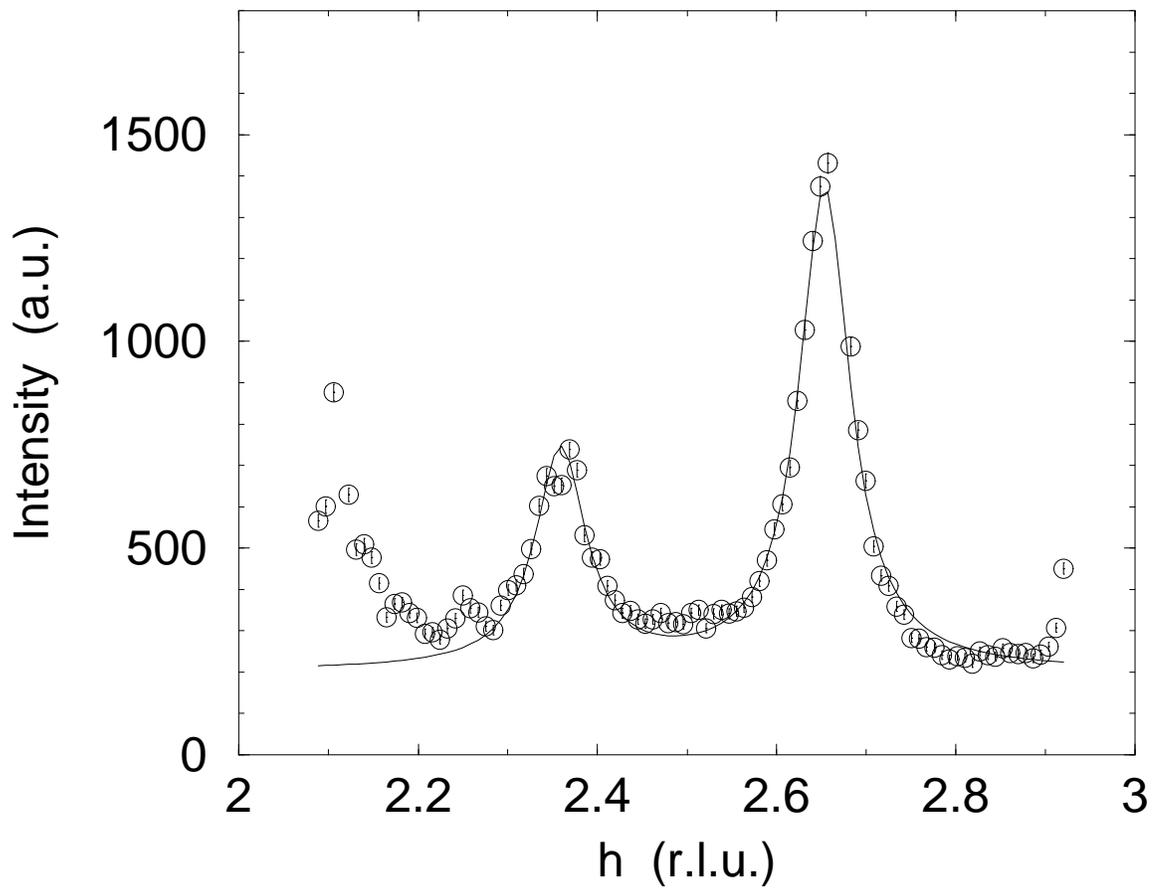
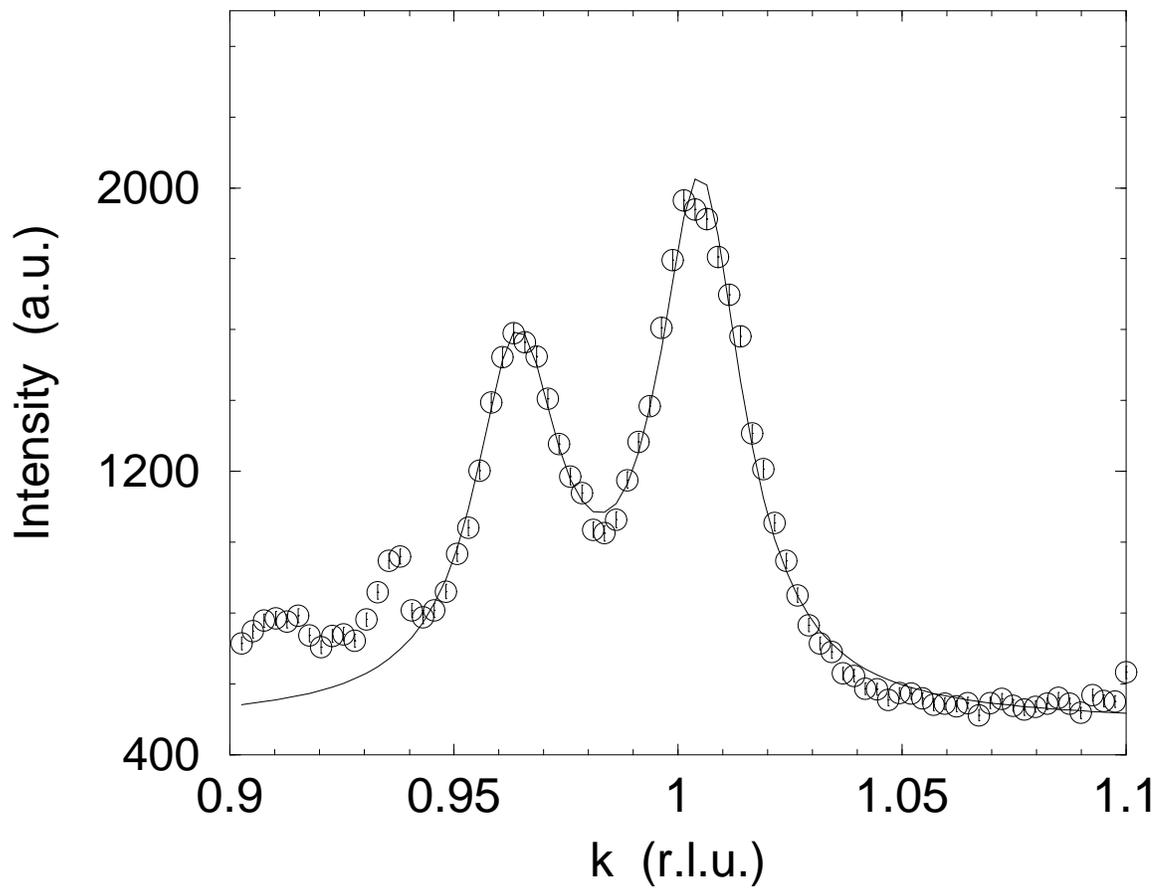
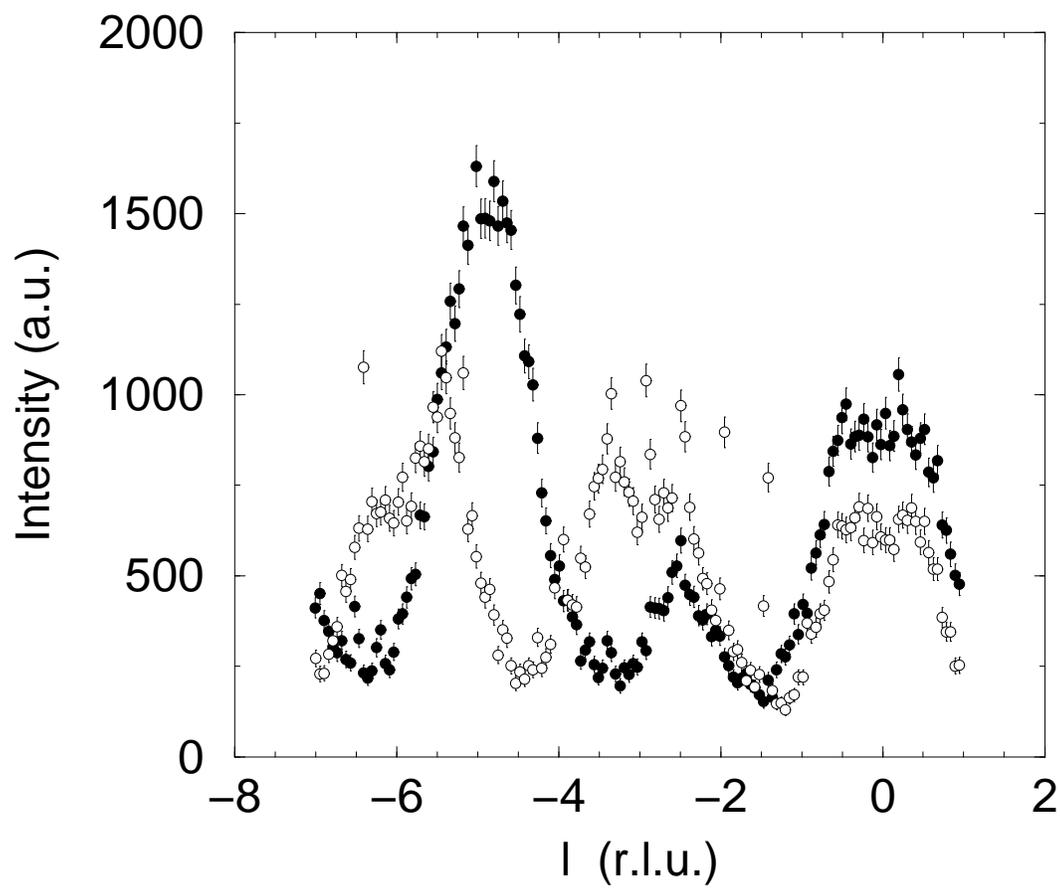
\begin{figure}
  \caption{Ortho-III superlattice reflections measured on a crystal with oxygen composition $x=0.72$ at room temperature.
  Left: $(h \; 0 \; 5)$-scan, middle: $(8/3 \; k \; 0)$-scan and right: scan along $(7/3 \; 0 \; l)$ (open circles) and
  $(8/3 \; 0 \; l)$ (filled circles). The lines shown with the $h$- and $k$-scan are fits to a Lorentzian function $(y'=1)$.}
  \label{ell_72}
\end{figure}

The ortho-III phase is found at the oxygen compositions of $x=0.72, 0.77$ and $0.82$. A
crystal prepared with $x=0.87$ showed no sign of any oxygen ordering. At these oxygen
compositions the ortho-III phase is formed by the sequence $(110)$ of two full chains and
one empty chain. Accordingly, the size of the unit cell is tripled along $a$ and the
diffraction pattern shows two superstructure reflections along $h$ between the
fundamental Bragg peaks. As shown in Fig. \ref{ell_72} the ortho-III super lattice peaks
are well defined in the $ab$ plane, but like all superstructures due to oxygen ordering
in YBCO, broadened due to finite domain sizes. In contrast to the ordering in the $ab$
plane, the $l$-dependence of the diffracted intensity shows only a broad modulation, with
a peak width corresponding to more than one reciprocal lattice unit. This $l$-modulation
is characteristic of the ortho-III structure and has been found in all samples exhibiting
the ortho-III phase\cite{Pla95,Pla92,Sch95a}, which indicates that the ordering of oxygen
atoms takes place in the $ab$-plane, whereas different planes are only weakly correlated.
Thus, in contrast to the ortho-II structure, which is 3D ordered, the ortho-III phase is
essentially a 2D ordered superstructure. As mentioned in Section \ref{Analysis} a simple
Lorentzian is expected for scattering from finite size domains in a 2D system when the
integration of the peaks is performed either along the $a$ or the $b$ direction
$(y'=y-1/2=1)$, whereas a Lorentzian to the power $y'=y=3/2$ is expected for integration
along $l$. However, the peak shapes along $h$ and $k$ are well described by Lorentzians,
independent of which component is integrated in the vertical direction. Attempts to
include a variable power $y$ did not improve the fits when the $l$ direction was
vertical. The smallest widths, which have been reported previously by Schleger {\it et
al.}\cite{Sch95a} are found in the $x=0.77$ crystal with $\Delta h = 0.031(1)$ and
$\Delta k = 0.0090(2)$.

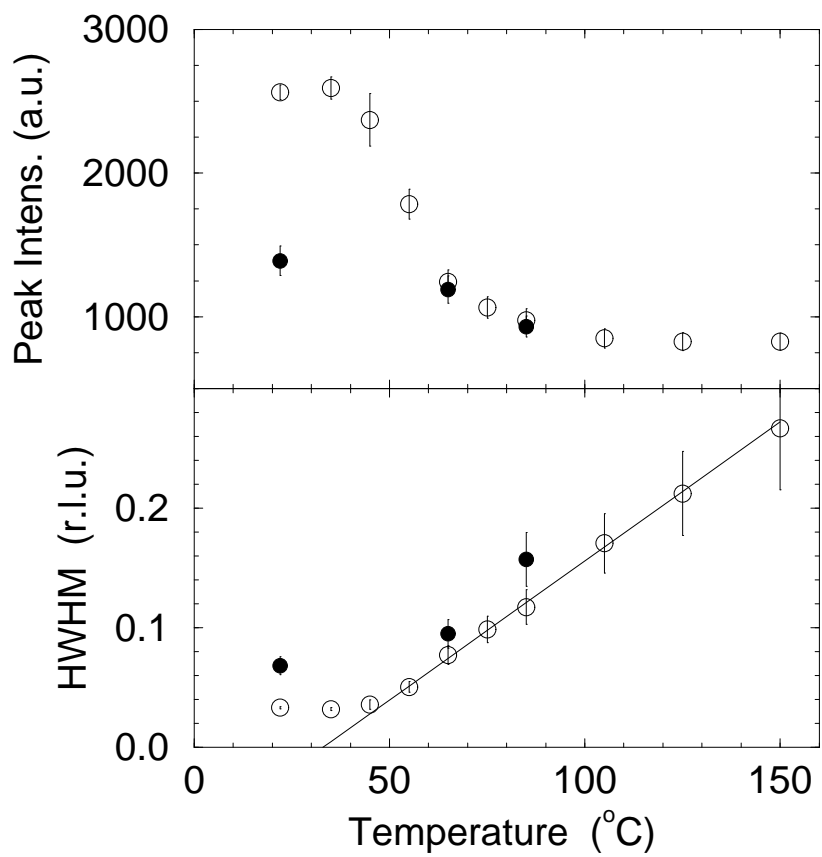
\begin{figure}
  \caption{Ortho-III transition at the oxygen composition $x=0.72$ where the peak intensity and the
    width of the $(8/3 \; 0 \; 5)$-reflection were measured by $h$-scans. The open circles show the heating
    data and the filled circles the data during cooling. The line in the lower graph shows the behavior
    described by the 2D Ising model.}
  \label{trans_72}
\end{figure}

One example of the temperature dependence of the ortho-III phase is shown in Fig.
\ref{trans_72} for the crystal with composition $x=0.72$. Here the $(8/3 \; 0 \; 5)$
reflection was scanned along $h$ at various temperatures. Similar to the transition of
the ortho-II phase the peak intensity and peak width are frozen at temperatures smaller
than 35$^{\circ}$C. Above this temperature critical fluctuations are observed. Fitting
the temperature dependence of the peak width a critical exponent of $\nu = 0.92(8)$ is
obtained, with a transition temperature of $T_{OIII}=48(5)^{\circ}$C. This value for the
critical exponent is in good agreement with the theoretical value of $\nu = 1$ for the 2D
Ising model and confirms the 2D character of the ordering.

\subsection{Ortho-V}
\label{ortho-V}

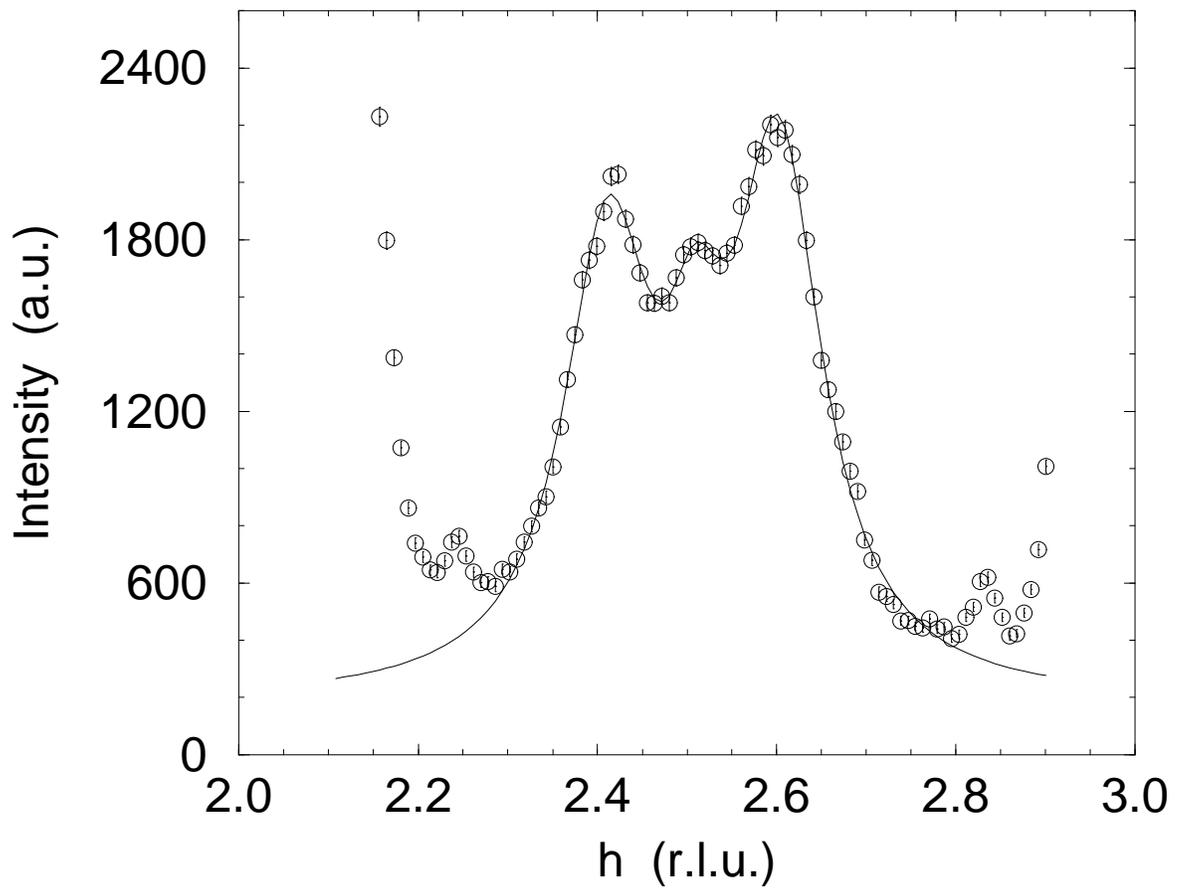
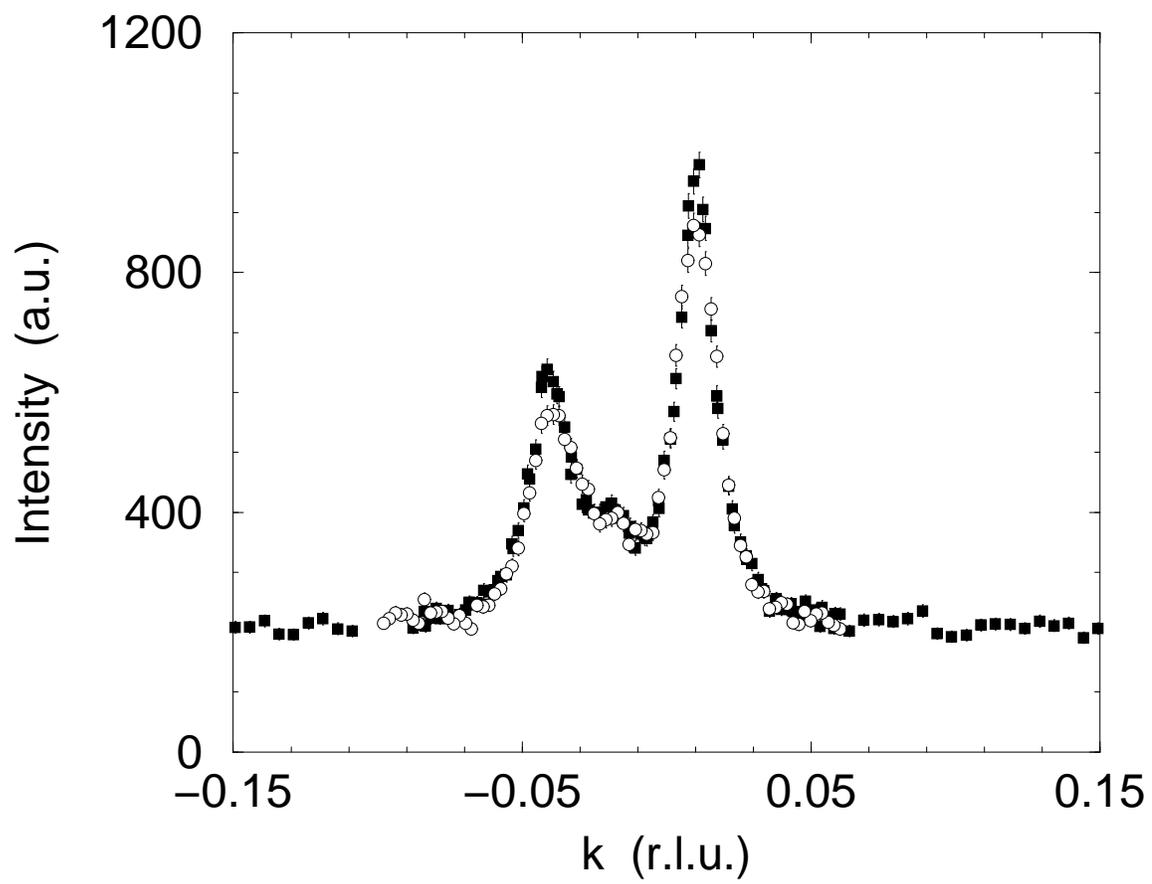
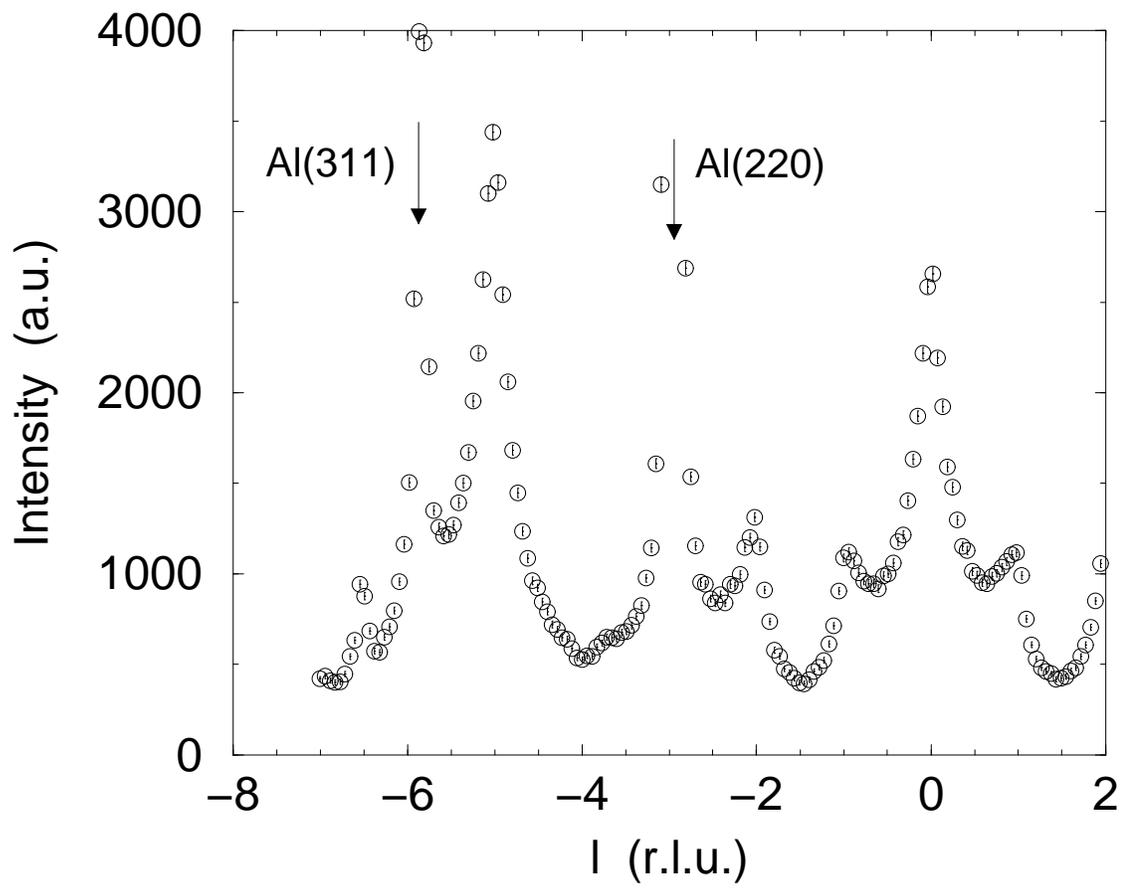
\begin{figure}
  \caption{Scans along $h$ (left), $k$ (middle) and $l$ (right) at room temperature for a crystal prepared to $x=0.62$.
  The peaks in the $(h \; 0 \; 0)$-scan at $h=2.4$ and $h=2.6$ result from the ortho-V phase.
  The scattering signal at $h=2.5$ indicates the presence of the ortho-II phase. The weak peak at $h=2.23$ is an Al-powder
  line originating from the sample holder and the bump at $h=2.83$ is unidentified. At $h=2.1$ and $h=2.9$ tails of the
  fundamental Bragg reflections are observed. In the middle part of the figure white circles show a scan along
  $(2.4 \;k \; 0)$ and black squares a $(2.5 \; k \;0)$-scan. The right hand side shows a $(2.5 \;0 \; l)$-scan. The arrows
  indicate the position of Al-powder lines.}
  \label{mix_hkl}
\end{figure}

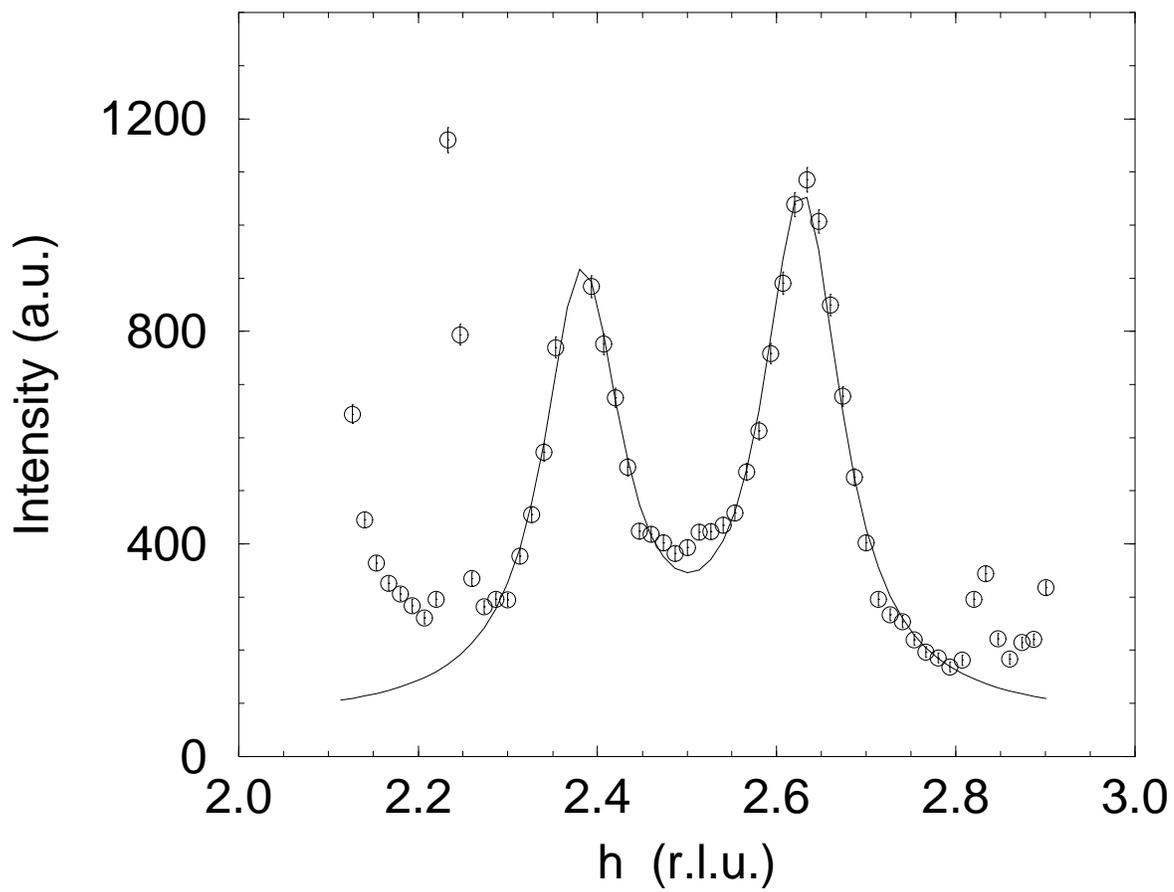
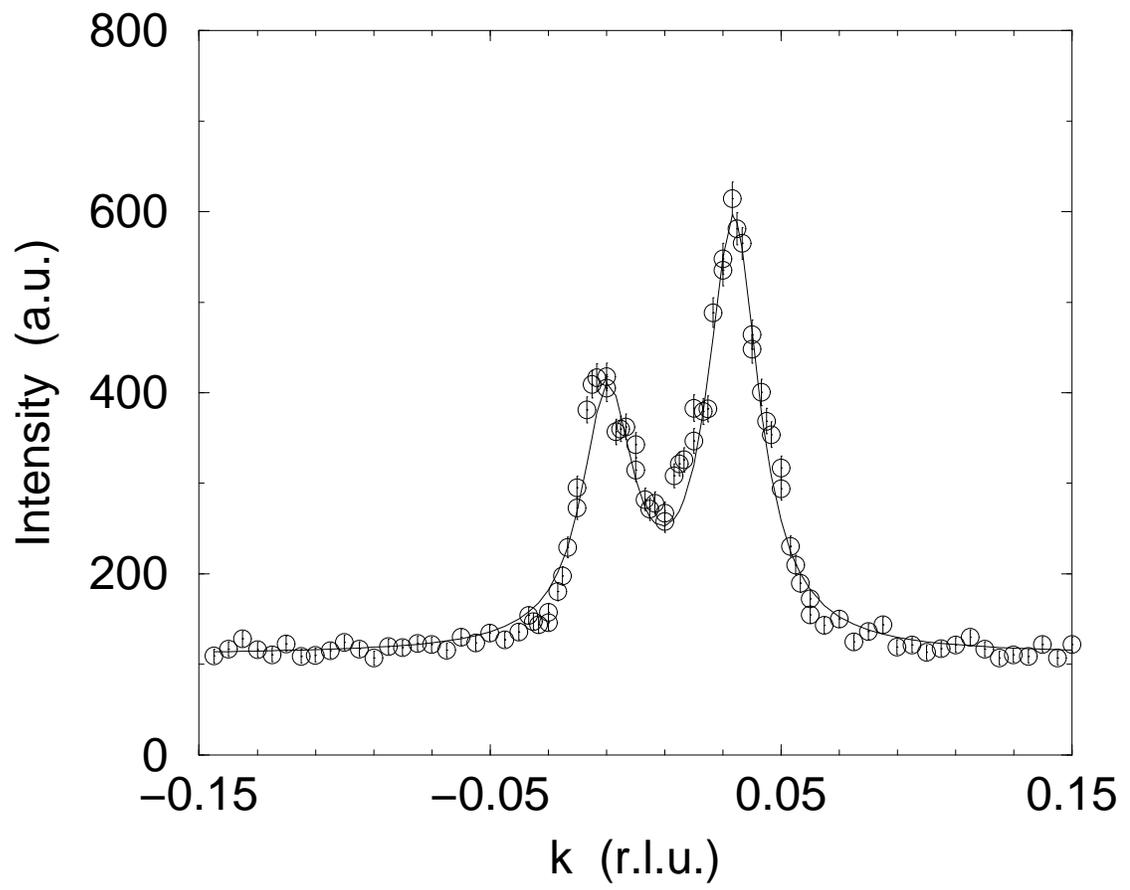
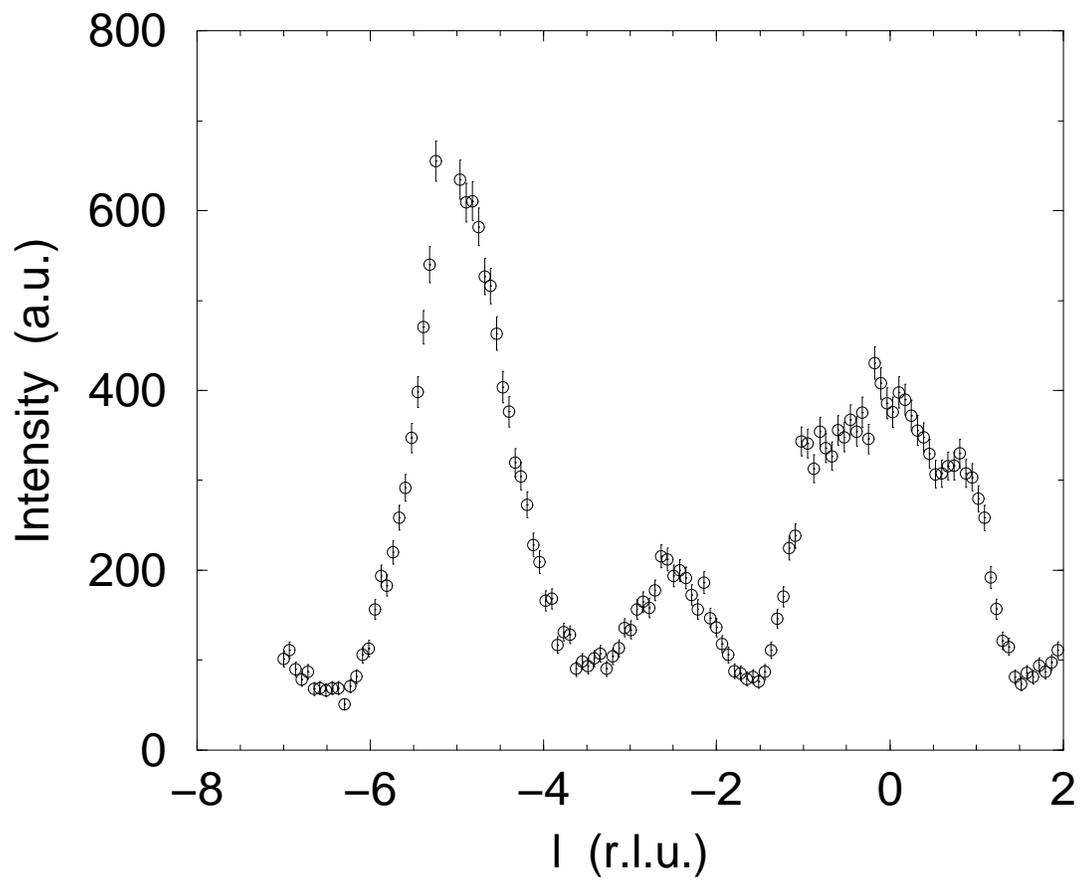
\begin{figure}
  \caption{Scan along $(h \; 0 \; 0)$ (left), $(2.63 \; k \; 0)$ (middle) and $(2.63 \; 0 \; l)$ (right) in a crystal with
  oxygen concentration of x=0.67 at room temperature. The $h$-scan on the left hand side shows peaks at approximately
  $h=2\frac{3}{8}$ and  $h=2\frac{5}{8}$ indicating the ortho-VIII phase. The peak at 2.23 is an Al-powder line.
   Lines shown together with the $h$- and $k$-scans are Lorentzian fits.  The variation of the
  diffracted intensity along $l$ is only weakly q-dependent and similar to that of the ortho-III phase.}
  \label{inco_hkl}
\end{figure}

The investigation of a crystal prepared with the oxygen composition of $x=0.62$ shows a
mixture of ortho-II and ortho-V phase at room temperature. This is revealed by the
observation of diffuse peaks at positions of $h=2.4, 2.5$ and $2.6$ as shown in Fig.
\ref{mix_hkl}. The peak at $h=2.5$ results from the ortho-II structure, and the peaks at
$h=2.4$ and $h=2.6$ are consistent with a unit cell which is enlarged five times in the
$a$ direction, {\it i.e.} the ortho-V structure. The two small peaks seen in the $h$-scan
in Fig. \ref{mix_hkl} at $h$=2.23 and $h$=2.83 are an Al-powder line and possibly a grain
of an unknown phase oriented with the lattice, respectively. The hump at $h$=2.83 has
also been observed when the same crystal was prepared with other oxygen stoichiometries
(compare with Fig. \ref{inco_hkl} and Ref. \ref{ref:Sch95a}). A similar diffraction
pattern, consistent with a mixture of ortho-II and ortho-V has been observed in all $(h
\; 0 \; l)$ scans performed with $1\le h\le 4$ and $l$=0,3,5,6,7 (8 scans in total).
However, none of these scans showed a peak at position $\vec{Q}=(\frac{1}{5} \; 0 \; l)$.
This is explained by the structure factor calculations of the superlattice peaks from the
ideal ortho-V ordering sequence (10110) shown in Ref. \ref{ref:deF90}. The intensities of
the peaks at $(1/5 \; 0 \; 0)$ are indeed much smaller than the ones at $(2/5 \; 0 \; 0)$
and $(3/5 \; 0 \; 0)$. However, this model takes into account only the oxygen order, and,
as discussed in Section \ref{intro}, the superlattice peaks are caused by both the oxygen
order and the cation displacements. These displacements and the pronounced disorder may
change the intensities and reduce them further.

Due to the heavy overlap of the peaks from the two phases it is difficult to determine
the peak shape and width. However, analysis of the ortho-II and ortho-V peaks using
Lorentzian profiles gave the following HWHM in reciprocal lattice units at room
temperature: $\Delta h = 0.040(28), \Delta k = 0.0078(16), \Delta l = 0.12(3)$ for
ortho-II, and $\Delta h = 0.058(10), \Delta k = 0.0096(19)$ for ortho-V. The scan along
$l$ at $(2.5 \; 0 \; l)$, shown in Fig. \ref{mix_hkl}, exhibits the intensity modulation
well known for 3D ordering in the pure ortho-II phase, but the heavy overlap of the peaks
prevents that the 2D short range type of modulation expected for the ortho-V peaks along
$l$ can be determined independently.

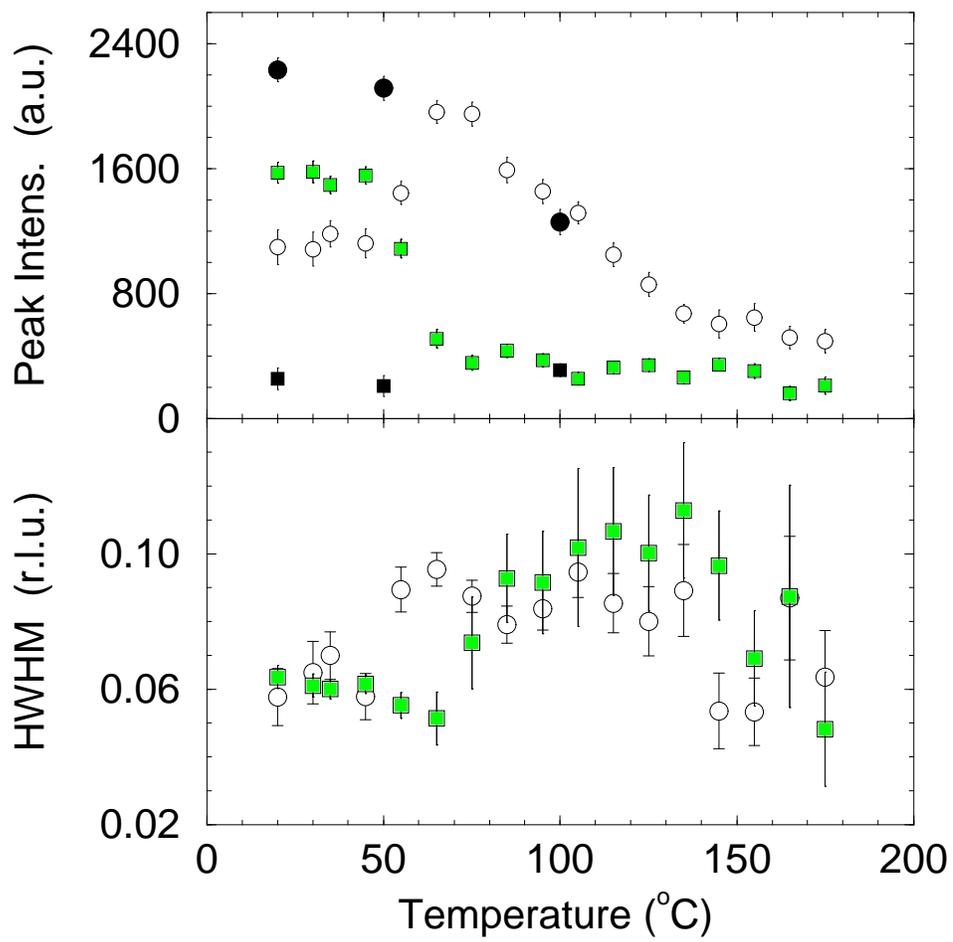
\begin{figure}
  \caption{Temperature dependence of the mixed phase of ortho-II and ortho-V at the oxygen composition of
  $x=0.62$. The top part plots the peak intensity of the ortho-II reflections (circles) and the ortho-V reflections
  (squares). Open and gray shaded symbols mark the data obtained during heating, the black filled symbols the cooling data.
  The bottom part of the figure shows the half width at half maximum (HWHM) of the reflections of the two phases along
  $h$.}
  \label{tempd_mix}
\end{figure}

The temperature dependence of this mixed phase was measured by $h$-scans between the $(2
\; 0 \; 0)$ and the $(3 \; 0 \; 0)$ Bragg reflections and the diffraction pattern was
fitted to three Lorentzians with fixed positions at 2.4, 2.5 and 2.6.  Looking at the
measurement of the phase transition of this mixed phase, shown in Fig. \ref{tempd_mix},
one observes that the ortho-V correlations disappear between 50$^{\circ}$C and
70$^{\circ}$C and at the same time the ortho-II gains intensity. Also during the cooling
cycle the ortho-II correlations dominate the diffraction pattern. The ortho-II
correlations disappear at approximately 110 $^\circ$C. Both facts together with our
knowledge about the ordering kinetics \cite{Sch95,Kal98} indicate that during the cooling
process the ortho-II phase is stabilized at higher temperature than ortho-V and with
rather fast ordering kinetics. Then at lower temperature the ortho-V phase becomes
stable, but due to the slow ordering kinetics at this lower temperature the ortho-V
domains do not form within the one hour time period of the experiment.

\subsection{Ortho-VIII}
\label{ortho-VIII}

Figure \ref{inco_hkl} shows $h, k$ and $l$ scans for the oxygen composition $x=0.67$. The
left part with $h$ scans along $(h \; 0 \; 0)$ with $2 < h < 3$ reveals diffuse
superlattice peaks at $h=2.382(4)$ and $h=2.627(3)$. The peak positions and profiles have
been fitted to two Lorentzians giving a HWHM  of $\Delta h=0.053$. The middle part shows
that the peaks are also localized in the transverse direction with a width of $\Delta
k=0.013(2)$. The modulation of the intensity for a scan along $l$ (right part of the
figure) has a similar $q$ dependence as the corresponding scan for the ortho-III phase,
shown in Fig. \ref{ell_72}. Thus, there are no well-defined peaks along $l$, indicating
essentially 2D ordering with substantial disorder in the stacking of full and empty
chains along the $c$ direction. Similar superlattice peaks have been observed at
positions in reciprocal space of $(h \; 0 \; 3)$ and $(h \; 0 \; 5)$ with $2\le h\le3$.
The superstructure peak positions at modulation vectors with $nh_m = 0.382$ and 0.627 are
close to the expected values $nh_m= 3/8$ and $5/8$ for a superlattice with a unit cell of
$8a \times b \times c$, {\it i.e.} the ortho-VIII phase. The expected sequence of full
and empty chains of the ideal ortho-VIII structure is (11010110). Calculating the
intensities of the superlattice peaks for this ideal case one finds that the observed
peaks at $nh_m=$ 3/8 and 5/8 are the strongest, the peaks at $nh_m =$ 2/8, 4/8 and 6/8
are about one order of magnitude smaller, and the ones at $nh_m=$ 1/8 and 7/8 are about
two orders of magnitude smaller (compare with the presentation in Ref. \ref{ref:deF90}).
Due to the weak ordering it is unlikely that the smaller superlattice reflections can be
observed.

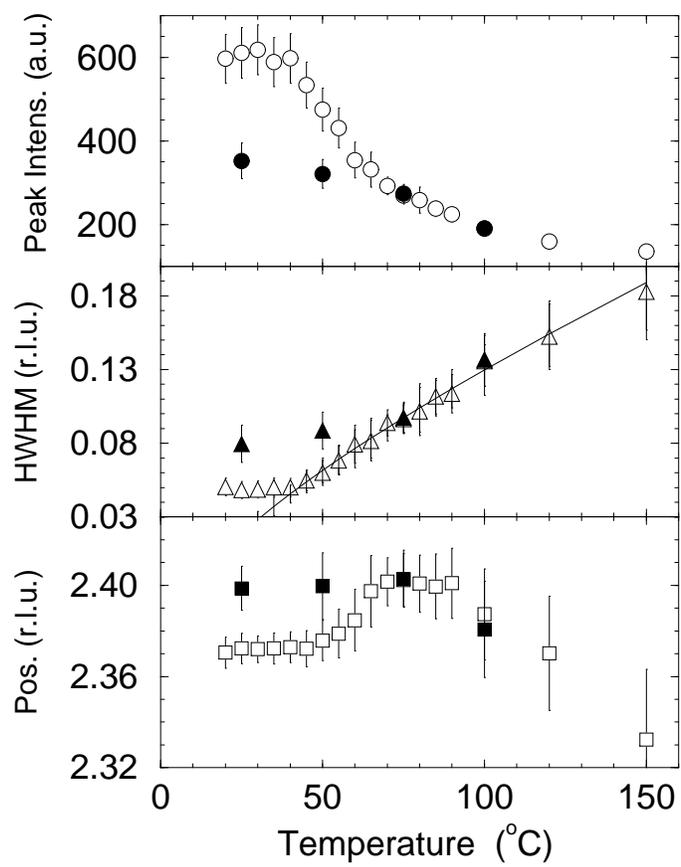
\begin{figure}
    \caption{\label{tempd_inco} Temperature dependence of the ortho-VIII phase at $x=0.67$ measured by
      $h$-scans on the $(2\frac{3}{8}\; 0\; 3)$ superstructure reflection. The top figure shows the peak
      intensity, the middle part the half width half maximum together with a fit to a critical behavior
      according to equation \ref{nu} with $\nu =0.79(3)$ and the bottom the peak position. The shift
      in the peak position indicates a transformation from ortho-VIII to ortho-V structure between
      50$^{\circ}$C and 70$^{\circ}$C. At 150$^{\circ}$C the peak position corresponds to the ortho-III
      structure.}
\end{figure}

The temperature dependence was observed from $h$-scans at the $(2\frac{3}{8} \; 0 \; 3)$
peak position and the results are shown in Fig. \ref{tempd_inco}. The onset of broadening
of the superstructure peaks takes place at $T_{OVIII}=42(5)^{\circ}$C. The temperature
dependence of the peak width above the transition temperature is described by the
critical exponent of $\nu =0.79(3)$ as shown is the middle part of Fig. \ref{tempd_inco}.
This value is between the exponent of 0.63 for the 2D and 1 for the 3D Ising model.
Another interesting feature of this phase transition is revealed by the inspection of the
peak position. When the temperature exceeds 50 $^{\circ}$C the peak position changes
continuously from $h=2.372$ to $h=2.4$ which corresponds to the position of the ortho-V
phase. Above 90$^{\circ}$C the peak shifts gradually to $h=2.33$ at 150$^{\circ}$C, the
location of the peaks of the ortho-III structure. Upon cooling the data are reproduced
down to 75$^{\circ}$C, at lower temperatures the intensity is significantly reduced and
the structure freezes into the ortho-V phase.

\subsection{The oxygen ordering phase diagram}
\label{Phase}

From the transition temperatures obtained in the present and previous
studies\cite{Sch95,Pou96,Sch95a} using hard x-ray diffraction and the same type of
crystals we may establish phase lines for the oxygen superstructure ordering. Combining
these data with the transitions temperatures, $T_{OI}$, of the phase transition from the
tetragonal to the orthorhombic ortho-I phase, obtained by neutron powder diffraction
\cite{And90}, we may construct the structural phase diagram of oxygen ordering in YBCO,
shown in Fig. \ref{phased}. In the figure is also included the phase transition
temperatures, $T_{OI}$ and $T_{OII}$ predicted by Monte Carlo simulations\cite{Fiig96}
based on the ASYNNNI model \cite{deF87} with {\it ab initio} interaction
parmeters\cite{Ste89}.

The only true equilibrium structures are the ortho-I phase and the tetragonal phase, all
superstructures formed by oxygen ordering do not show long range ordering. Within the
temperature range studied the tetragonal phase is the only one observed for $x
< 0.35$. Below the tetragonal to orthorhombic phase transition temperature the 3D ordered ortho-I phase always develops,
and it is the only structure observed for $x > 0.82$. For $ 0.35 \le  x < 0.62$ the 3D
short range ordered ortho-II phase is the only stable superstructure. Similarly, a single
phase ortho-III structure with 2D finite size ordering is observed for $0.72 \le x \le
0.82$. At intermediate compositions a mixed phase of ortho-II and ortho-V is found at
$x=0.62$, and ortho-VIII is found at $x=0.67$ in crystals that have been slowly cooled to
room temperature as described in Sec. \ref{Sample Preparation}. Both the ortho-V and the
ortho-VIII structures are essentially 2D ordered and have finite size ordering. During
heating the ortho-V structure transforms into ortho-II and it does not recover on cooling
within one hour. Above room temperature the ortho-VIII structure transforms gradually
first into ortho-V and then into ortho-III. On subsequent cooling the ortho-V
superstructure is recovered and remains stable within the one hour time period of the
measurements.

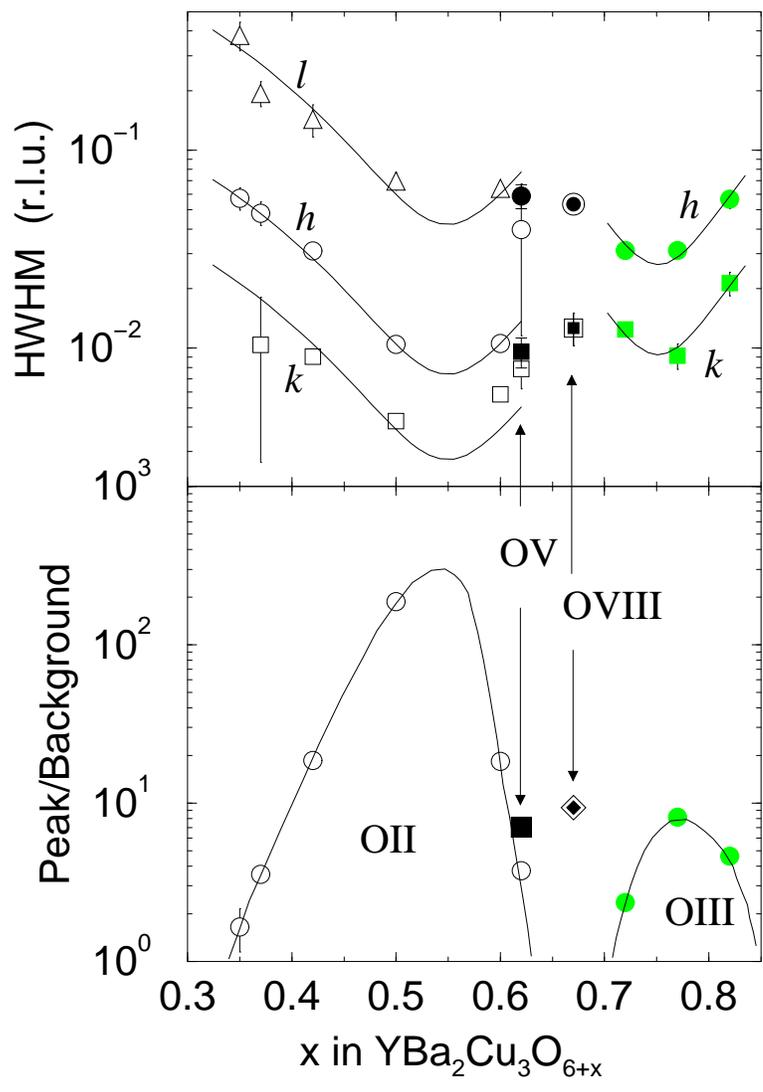
\begin{figure}
  \caption{Top: Half width half maximum for the indicated structures as function of
  oxygen content. Bottom: peak to background ratio. All lines are guides to the eye.}
  \label{res_pap}
\end{figure}

The line shape of the superstructure reflections is in most cases well described by a
simple Lorentzian ($y=1$). Only for the ortho-II phase between $0.42\le x < 0.62$ a
Lorentzian squared shape ($y=2$) is found. At the low oxygen side of the ortho-II phase
$x\le 0.36$ the small peak to background ratio (see bottom part of Fig. \ref{res_pap})
does not permit the determination of the exponent of the Lorentzian. The domain size of
the superstructures depends strongly on the crystal quality and the annealing times.
However, for high quality crystals that have been annealed by the standard procedure for
studies of the phase diagram (described in Sec. \ref{Sample Preparation}) we expect that
the domains are at the late state of growth and therefore only weakly time dependent
({\it cf.} Fig. \ref{peak-HWHM-prep}, thermal preparations 1 and 2, and Refs.
\ref{ref:Sch95} and \ref{ref:Kal98}). On this basis we consider the results presented in
Sec. \ref{Results} of the peak widths measured at room temperature after the initial
thermal preparation as saturation values. The HWHM of the superlattice peaks measured
along the three axis of reciprocal space as function of oxygen composition is depicted in
Fig. \ref{res_pap} (top). The parallel lines (guides to the eye) in the logarithmic plot
observed in the ortho-II phase as well as in the ortho-III phase show that the ratio of
the anisotropy is constant within a given structural phase. For the ortho-II phase we
find the following ratios of the inverse correlation lengths at room temperature:
$\Gamma_h/\Gamma_k=2.7(6)$ and $\Gamma_l/\Gamma_k=15(2)$. The $ab$ plane ratio seems to
be independent also of the type of structure, since the ratio for the ortho-III phase:
$\Gamma_h/\Gamma_k=2.9(4)$ is in good agreement with the value of the ortho-II phase.
This implies that the domain pattern in the $ab$ plane scales in both the ortho-II and
the ortho-III phases, and in ortho-II the scaling is extended to 3D.  The peak
intensities cannot be compared directly because different crystals and instrumental
settings have been used. However, the peak intensity normalized to the background, shown
in the bottom part of Fig. \ref{res_pap}, is essentially an independent parameter of the
ordering properties. From this normalized peak intensity and the HWHM data it is clear
that the optimal superstructure order parameter is found close to $x = 0.55$.

The oxygen composition $x$ of all the ordered phases deviates systematically from the
ideal composition of these phases. For example the longest correlation lengths for the
ortho-II phase are likely to be at $x\simeq 0.55$. Unfortunately no data points are
available at this composition. Theoretically one would expect the best ortho-II order for
$x=0.50$. This deviation is even more significant for the ortho-III phase, which is
expected at $x=0.67$, but observed around $x\simeq 0.77$. Thus, the deviation from the
ideal composition increases with increasing chain density and the amount of oxygen atoms
occupying sites on the empty chains at room temperature can be estimated to be about 10
\% for ortho-II and 30 \% for ortho-III.

\section{Discussion}
\label{Discussion}
\subsection{Experimental results}

It has been known for several years that the ortho-II and ortho-III superstructures are
bulk structural phases of finite size domains. Several other superstructures have been
suggested, mainly from electron microscopy. In the present paper we have shown that also
the ortho-V and ortho-VIII correlations observed by electron microscopy result from bulk
structural ordering, but we found no evidence for the ortho-IV phase. However, we
recognize in particular the early electron microscopy results obtained by Beyers {\it et
al.}\cite{Bey89} which are in close agreement with our room temperature data. Beyers {\it
et al.} observe the ortho-II and ortho-III superstructures in the same composition range
as in our studies. Furthermore, they found co-existence of ortho-II and ortho-V  at
$x=0.65$, and a structure similar to the ortho-VIII phase, which they call a '$(0.37 \; 0
\; 0)$' structure, at $x=0.71$.

Beyers {\it et al.} attributed the clear disagreement between the observed oxygen
compositions and the stoichiometries of the ideal superstructure phases to gradients in
the oxygen content of the sample, which might be different on the surface and in the bulk
material. In our experiment such differences can be ruled out. We conclude that this
deviation is an intrinsic property of the oxygen ordering mechanism. It is possible that
the phase lines between the superstructure phases  are in fact tilted, and only at zero
temperature the ideal oxygen stoichiometry of the superstructure phases is found.
However, this will never happen because the oxygen ordering kinetics is very slow at the
temperatures where the superstructures become stable, and the movement of Cu-O chains
freeze effectively below approximately 40 $^\circ$C.

Beyers {\it et al.} interpret the mixing of ortho-II and ortho-V phases at $x=0.65$ as a
phase separation, which leads to the 60 K plateau \cite{Bey89}. Our investigation of the
temperature dependence together with the studies of the ordering kinetics
\cite{Sch95,Kal98} may lead to a different conclusion. During the cooling of a sample
with an oxygen content of $x=0.62$ (in the case of Beyers {\it et al.} $x=0.65$) oxygen
starts to order in the ortho-II phase. The relatively high temperature enables a fast
growth of ortho-II domains. At lower temperature the ortho-V phase becomes stable, but
now at temperatures just above room temperature where the growth of ortho-V domains is
slow. A full transformation into the ortho-V phase cannot be precluded but it is very
time consuming. Therefore, we suggest that domains of the complex ortho-V superstructure
start to grow inside the ortho-II structure and a mixed phase, rather than phase
separation, results.

From studies of the oxygen ordering properties it has become clear that the finite size
of the ortho-II superstructure results from formation of anti-phase boundaries that limit
the domain growth due to slow kinetics of moving long Cu-O chains. The reason for this
has been discussed by Schleger {\it et al.} \cite{Sch95}, and it was speculated that
random fields introduced by impurity defects in the crystal stabilize the anti-phase
domain walls and prevent formation of long range order. This is corroborated by the
present studies and additional studies of the ordering kinetics \cite{Kal98}. However,
the observation of superstructures extending over eight unit cells shows the importance
of long range interactions for the ordering mechanism. That these long range interactions
play a significant role for the finite size ordering has recently been established by
model simulations \cite{Mon98}, and will be discussed further below. For the ortho-III,
ortho-V and ortho-VIII superstructures the small 2D domains indicate that the ordering
resembles a random faulting sequence of ortho-II and ortho-III. Khachaturyan and Morris
\cite{Kha90} have suggested that this is a likely ordering scheme, and they have
calculated structure factors which are qualitatively similar to those observed at room
temperature in our experiments. However, the fact that the ortho-V and ortho-VIII
superstructures only appear when they are slowly cooled indicates that the long range
interactions tending to form these superstructures become effective at low temperatures,
but the slow oxygen ordering kinetics for movement of long Cu-O chains prevent that
well-defined domains are formed. As mentioned in Sec. \ref{Analysis} we would expect a
diffraction profile of a Lorentzian to the power $y'=y=3/2$ from 2D domains with sharp
boundaries when the integration along the $c$ axis is incomplete. The observation that
all the superstructure peaks of the ortho-V, ortho-VIII and ortho-III peaks are described
by Lorentzian profiles suggests that these superstructures have a more fuzzy type of
boundaries than the ortho-II domains.

Generally, there is significant hysteresis in the superstructure ordering when the
temperature is cycled through the phase transitions.  The ortho-II and ortho-III
superstructures are re-established during cooling from the ortho-I phase within one hour.
However, the ortho-V phase (mixed with ortho-II) and the ortho-VIII do not recover during
cooling within this short time period. Instead, the less complex superstructures,
ortho-II and ortho-V develop, respectively.  It is obvious that the superstructure
ordering does not represent equilibrium phases, and it cannot be precluded that more
complex superstructures may be formed by very long time annealing at an appropriate
temperature or in crystals that are even more perfect than the present ones.
According to Ostwald's step rule for phase transformations, metastable phases
may be formed, before the system finally transforms into the stable phase, as long as 
nucleation centers with a similar structure like the metastable phases are present.
In our case the ortho-II and ortho-III phase might be nucleation centers for the 
ortho-V phase, which in turn, at $x=0.67$, is metastable and serves as a nucleation 
center for the ortho-VIII phase (see figure \ref{tempd_inco}). Thus,
although we have been able to define unique transition temperatures, at least for the
ortho-II superstructure phase, it is questionable whether we have established a phase
diagram in the usual sense. This may explain why the phase diagram does not comply with
Gibb's phase rule.

The 3D ordered superstructures with unit cells $2 \sqrt{2}\, a \times 2 \sqrt{2}\, a
\times c$ and $\sqrt{2} a \times 2 \sqrt{2} a \times c$, the so called herringbone
structure, have been observed by electron microscopy, and one group has reported on these
structures by neutron \cite{Son91} and x-ray \cite{Zei92a} diffraction techniques on
single crystals with composition $x = 0.35$. However, no other experiments with bulk
structural techniques, could confirm these results. Bertinotti {\it et al.} \cite{Ber89}
and Yakhou {\it et al.} \cite{Yak96} have shown that the reflections of the herringbone
type can be assigned to BaCu$_3$O$_4$ grains in the crystals. Krekels {\it et al.}
\cite{Kre90,Kre95} attribute the $2 \sqrt{2}\, a \times 2 \sqrt{2}\, a \times c$
structure to distortions of the CuO$_5$ pyramids in the CuO$_2$ planes, and Werder {\it
et al.} \cite{Wer91} suggest that they could result from ordering of copper and
barium vacancies in the lattice. The consensus from these and several other studies is
therefore that the $2 \sqrt{2}\, a \times 2 \sqrt{2}\, a \times c$ and the herringbone
type structures are not oxygen ordering superstructures in YBCO. If they were, it is
peculiar that they have 3D long range order while the Cu-O chain ordering develops only
finite size domains. Also, we have found no evidence of them at any composition $x$ in
the present hard x-ray diffraction studies on carefully prepared high quality single
crystals.

\subsection{Significance for superconductivity}
The significance of the oxygen ordering for charge transfer and superconductivity is
obvious from many studies. Chemical bond considerations combined with structural
\cite{Cava90} and spectroscopic studies \cite{Tol92,Lue96} have shown that the basal
plane copper in undoped YBa$_2$Cu$_3$O$_{6+x}$, $(x=0)$, is monovalent and that simple
oxygen monomers, {\it i.e.} Cu-O-Cu, will not give rise to charge transfer. However,
charge transfer is observed for larger $x$ where Cu-O chains are formed. Cava {\it et
al.} \cite{Cava90} and Tolentino {\it et al.} \cite{Tol92} have established that an
increasing amount of oxygen give rise to a charge transfer to the CuO$_2$ planes that is
in good agreement with the well-known plateau variation of $T_c$ with $T_c=58$ K around
$x=0.5$ and $T_c=93$ K close to $x=1$. Relating the oxygen ordering to the variation of
$T_c$ observed {\it e.g.} by Cava {\it et al.} \cite{Cava90} we find that the 58 K
plateau is identical to the stability range of the ortho-II superstructure, the rise in
$T_c$ from the 58 K to the 93 K plateau takes place at values of $x$ where the
ortho-V/II, ortho-VIII and ortho-III structures are found, and the 93 K plateau coincide
with the oxygen compositions of the ortho-I phase.

The significance of the ortho-II ordering for superconductivity has been shown directly
by Veal {\it et al.} \cite{Vea90} and Madsen {\it et al.} \cite{Mad97}. Both groups have
shown that $T_c$ is significantly reduced just after a quench and increases with time
towards the equilibrium values with a thermal activated time constant. The conclusion,
that may be drawn from these experiments and the present structural data, is that the
formation of ortho-II superstructure is decisive for the charge transfer and $T_c$. When
the sample is quenched from temperatures above $T_{OII}$, and even from the tetragonal
phase, it is only the time used to quench it into the ortho-II phase that matters. A time
dependent increase of $T_c$ is observed at annealing temperatures down to 250 K. Most
likely this temperature is the lower limit for local oxygen jumps which dominates the
oxygen ordering at the very early time. It is unlikely that the domain wall separating
anti-phase domains are mobile at 250 K.

\subsection{Relation to model calculations}
There has been many theoretical studies of the oxygen ordering in YBCO and attempts to
correlate the structural ordering with the electronic properties and superconductivity.
These include phenomenological relations between ordered oxygen domains and $T_c$
\cite{Pou91,Pou91a}, and electron band structures calculated from oxygen chain
configurations estimated {\it ad hoc}\cite{Zaa88} or derived from model studies
\cite{McC92}. More realistic and elaborate models, where the electronic degrees of
freedom from the strongly correlated electron system has been included in combination
with the oxygen ordering properties, have also been considered \cite{Uim94,Ali94,Hau97}.
The aim is clearly to understand details of the local oxygen ordering properties which
are important for the electronic structure and the charge transfer, but difficult to
obtain directly from experiments. The predictive power of these model studies for the
structural and electronic properties is strongly related to their ability to reproduce
the experimental findings, as presented in the present structural studies.

Most of the structural models are based on local effective oxygen-oxygen interactions in
a 2D lattice gas formulation \cite{deF87,Ali90}. These models do not take into account
long range interactions like strain effects and will therefore not reproduce the twin
domain formation observed experimentally at the onset of the ortho-I ordering. Models
including such long range interactions have been considered \cite{Sem92}. It has also
been suggested that the diffuse scattering results from generation of $(1\ 0\ 0)$
interstitial plane defects that order by forming a Magneli type homologous series
\cite{Kha88}. However, the oxygen ordering observed experimentally has a predominant 2D
character related to the CuO$_x$ basal plane, that is $(0\ 0\ 1)$ planes. It takes place
inside the ortho-I twin domains and has domain sizes that are usually much smaller than
the twin domains. Strain effects and long range Coulomb type interactions are therefore
of little significance for the oxygen superstructure formation but they do play a role
for the mesoscopic ordering properties.

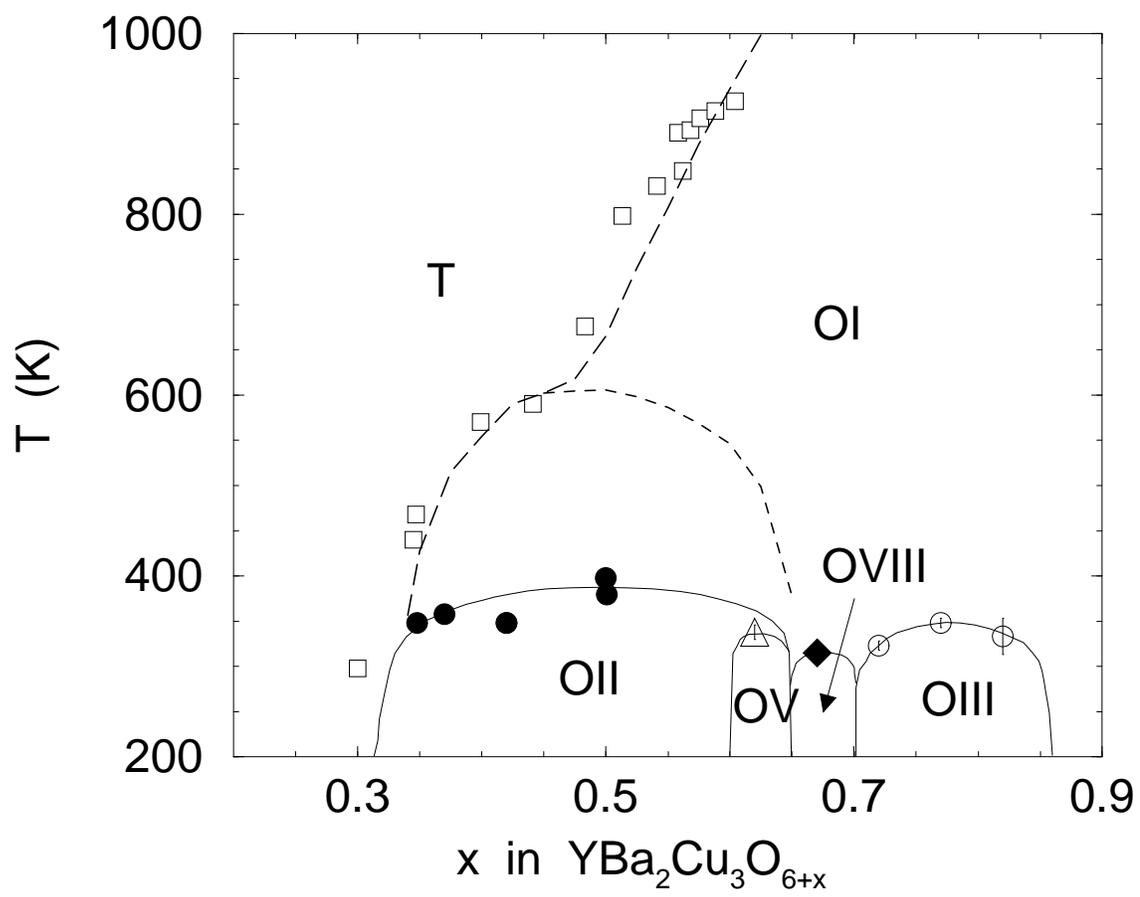
\begin{figure}
  \caption{The structural phase diagram of YBCO. The structural phases and their transition temperatures are labeled:
   teragonal (T), ortho-I (OI,$\Box$), ortho-II (OII,$\bullet$), ortho-III (OIII,$\circ$), ortho-V (OV,$\triangle$)
   and ortho-VIII (OVIII,$\diamond$). Solid lines are guides to the eye. The dashed lines are predictions from the
   ASYNNNI model. The data for the T-OI transition ($\Box$) are from Andersen {\it et al.} \cite{And90}. The
   $T_{OII}$ transition temperatures for $x$ = 0.35 and $x$= 0.36 are from Poulsen {\it et al.} \cite{Pou96}, the upper
   data set for  $x$=0.50 is from Schleger {\it et al.} \cite{Sch95}, and the $T_{OIII}$ transition temperature at $x$=0.77
   is from Schleger {\it et al.} \cite{Sch95a}.}
  \label{phased}
\end{figure}

The simplest model, that accounts for many elements of the oxygen ordering properties,
like the formation of Cu-O chains and the presence of the tetragonal, ortho-I and
ortho-II phases, is the so-called ASYNNNI (Asymmetric Next Nearest Neighbor Ising) model
\cite{deF87}. The ASYNNNI model is a 2D lattice gas (or Ising) model with effective
oxygen-oxygen pair interactions, that are assumed to be independent of temperature and
composition $x$. The interactions parameters include a strong Coulomb repulsion $V_1$
between the oxygen atoms on nearest neighbor sites and an attractive covalent interaction
$V_2$ between oxygen atoms that are bridged by a Cu atom. These two interactions locate
the oxygen on the $b$ axis (the O(1) site) and prevent significant oxygen occupation on
the $a$ axis (the O(5) site) in the orthorhombic phases at moderate temperatures and
compositions $x>0.35$. A weaker effective repulsive Coulomb type interaction $V_3$
between oxygen atoms that are next nearest neighbors and not bridged by a Cu atom
stabilizes the ortho-II superstructure. The ASYNNNI model accounts quantitatively
\cite{Fiig96} for the temperature and composition dependence of the experimental
structural phase transition between the tetragonal and the ortho-I phases \cite{And90}
(see Fig. \ref{phased}) by use of interaction parameters, which are consistent with values
obtained by Sterne and Wille \cite{Ste89} from first principles total energy
calculations: $V_1/{k_B} = 4278$ K, $V_2/{k_B} = 1488$ K and $V_3/{k_B} = 682$ K. It also
predicts the existence of the ortho-II phase, but it cannot account for the additional
superstructure phases, ortho-III, ortho-V and ortho-VIII. Moreover, it predicts long
range order of the superstructure phases, which has never been obtained experimentally,
and it cannot account quantitatively for the ortho-I to ortho-II phase transition
temperature.

Extensions of the ASYNNNI model have been suggested to account for the shortcomings.
These include an effective 3D interaction with a nearest neighbor attractive interaction
along $c$ that is $V_4 \approx - 0.02 V_1$ \cite{Fiig96}, effects of electronic degrees
of freedom in the Cu-O chain structure, as mentioned above \cite{Uim94,Ali94,Hau97}, and
2D Coulomb type interactions of longer range than $V_2$ and $V_3$ \cite{deF90,Ali94a}.
For the 2D ordering it has been argued that a single additional interaction parameter for
oxygen atoms that are $2a$ apart and not bridged by copper should be sufficient
\cite{Mon98}. This is corroborated by an estimate of screened Coulomb potentials which
shows that the interaction between oxygen atoms separated by $2a$ is of the order of $V_5
= 0.02 V_1$, and it decays rapidly for larger distances. At the temperature where $V_5$
becomes effective Cu-O chains have already been formed \cite{Mon98} and it will act as an
effective interaction between chains rather than between oxygen pairs. The $V_5$
interaction stabilizes the ortho-III phase by construction but it is not expected to
account for the ortho-V and ortho-VIII phases. The influence of including effective
Coulomb type Cu-O chain interactions extending beyond $a$ and $2 a$ has been studied
analytically in the framework of a 1D Ising model \cite{deF90}. Here a sequence of
branching phases develops for $T \rightarrow 0$ in order to comply with the Nernst
principle and stoichiometric phases at different compositions $x$. However, for the YBCO
system it is expected that the interactions of range beyond $2 a$ play a role only at low
temperatures where effectively the structural ordering is frozen. Also, the projection to
a 1D system requires that rather long Cu-O chains are formed, and recent Monte Carlo
simulations have shown that finite chain lengths may result when the ASYNNNI model is
extended by the $V_5$ interaction, even for $T \rightarrow 0$ \cite{Mon98}. On the other
hand, the ASYNNNI model extended this way predicts that the $V_5$ parameter is sufficient
to establish not only short range correlations of ortho-V and ortho-VIII but also the
ortho-II and ortho-III superstructures do not develop long range order, as observed
experimentally. The finite size ordering of the superstructures is therefore not
necessarily a consequence of impurities or defects that pin the domain walls, but may be
an intrinsic disordering property. Experiments on even more perfect crystals than used in
the present study could supply additional information about the influence of defects on
the ordering properties. In further agreement with the experiments, the ASYNNNI model
extended with the $V_5$ as well as the 3D $V_4$ parameters predicts a significant
suppression of the $T_{OII}$ ordering temperature relative to the $T_{OI}$ temperature
which the original version failed to do (see Fig. \ref{phased}). It is therefore, a
promising model for analysis of experimental results and credible predictions about the
local oxygen ordering properties. So far the theoretical phase diagram including the
$V_4$ and $V_5$ interactions has not been determined. The present experimental results
can be used as a guide to further model studies. Here it is interesting to note that our
data show that the ratios of the correlation lengths are essentially independent of
the oxygen stoichiometry in the ortho-II and the ortho-III phases. A comparison between
this result and mean field predictions of the peak widths \cite{Fii95} indicates that the
ASYNNNI model interaction parameters are independent of $x$. This assumption has been a
major objection against the validity of the ASYNNNI model.

\section{Concluding summary}
\label{Summary}

High energy X-ray diffraction has proven to be a unique tool for studies of oxygen
ordering properties in the orthorhombic phase of YBCO. Chain ordered superstructures of
the ortho-II, ortho-III, ortho-V and ortho-VIII types have been observed in high quality
single crystals with this bulk sensitive technique. None of the superstructures develops
long range order. Only the ortho-II phase is a 3D ordered superstructure with anisotropic
correlation lengths. The ortho-II correlation lengths observed at room temperature depend
on the oxygen composition (optimal for $x=0.55$), crystal perfection and thermal
annealing. All other superstructures have 2D character with ordering only in the
$ab$-plane. The ratio of the $ab$ plane correlation lengths is essentially independent of
the oxygen composition and whether the ordering is ortho-II or ortho-III. The transition
temperatures of the superstructures are between room temperature and 125$^{\circ}$C. The
ordering properties resulting from thermal cycling through the $T_{OII}$ and the
$T_{OIII}$ ordering temperatures show that finite size domains with internal
thermodynamic equilibrium are formed. The domain size observed on cooling from the
ortho-I phase within one hour is significantly reduced compared to the value obtained by
long time annealing. The observation of ortho-V mixed with ortho-II, and ortho-VIII
superstructures shows that these superstructures are bulk properties, and that Coulomb
interactions beyond next-nearest-neighbors become effective close to room temperature.
The ordering of the ortho-V and ortho-VIII superstructures does not reproduce when the
sample is cooled from the ortho-I phase within one hour, and it can not be precluded that
additional superstructure phases may be formed by careful annealing of high quality
single crystals. Therefore, although an unambiguous criterion has been identified for the
ordering temperatures of the finite size ortho-II and ortho-III superstructures, the
resulting 'phase diagram' is not an equilibrium phase diagram in the usual sense.

\section*{Acknowledgments}

This work was supported by the EC TMR - Access to Large Scale Facilities Programme at
HASYLAB, and the Danish Natural Science Research Council through DanSync. The Danish
Technical Science Research Council supports TF. Collaboration with H. Casalta, R.
Hadfield and P. Schleger on initial studies preceding this work is gratefully
acknowledged. Technical assistance from S. Nielsen, R. Novak, A. Swiderski and
T. Kracht is much appreciated.


\begin{thebibliography}{}

\bibitem[*]{MvZ}
    present address: Deptartment of Physics, Brookhaven National Laboratory, Upton, New York 11973, USA
 
\bibitem{And90}
    N.H. Andersen, B. Lebech and H.F. Poulsen, Physica C \textbf{172}, 31 (1990).

\bibitem{Sch91}
    P. Schleger, W. Hardy and B. Yang, Physica C \textbf{176}, 261 (1991).

\bibitem{Jorg87}
    J.D. Jorgensen, M.A. Beno, D.G. Hinks, L. Soderholm, K.J. Volin, R.L. Ritterman, D.G.
    Grace, I.K. Schuller, C.U. Segre, K.Z. Zhang and M.S. Kleefisch, Phys. Rev. B \textbf{36},
    3608 (1987).

\bibitem{Cla90}
    H. Claus, S. Yang, A.P. Paulikas, J.W. Downey and B.W. Veal, Physica C \textbf{171}, 205
    (1990).

\bibitem{Vea90}
    B.W. Veal, A.P. Paulikas, H. You, H. Shi, Y. Fang and J.W. Downey, Phys. Rev. B \textbf{42},
    6305 (1990).

\bibitem{Lib93}
    S. Libbrecht, E. Osquiguil, B. Wuyts, M. Maenhoudt, Z.X. Gao and Y. Bruynseraede, Physica C
    \textbf{206}, 51 (1993).

\bibitem{Mad97}
    J. Madsen, N.H. Andersen, M. v. Zimmermann and Th. Wolf, Annual progress 
report of the Department of Solid State Physics 1 January - 31 December 1996, 
page 47 (1997). Risø Report R-933 (EN).

\bibitem{Sch95}
   P. Schleger, R. Hadfield, H. Casalta, N.H. Andersen, H.F. Poulsen, M. von Zimmermann, J.R. Schneider,
   Ruixing Liang, P. Dosanjh and W.N. Hardy,
   Phys. Rev. Lett. \textbf{74}, 1446 (1995).
    \label{ref:Sch95}

\bibitem{Zan87}
   H.W. Zandbergen, G. van Tendeloo, T. Okabe and S. Amelinckx, Phys. Stat. Sol. \textbf{103}, 45 (1987).

\bibitem{Rey89}
   J. Reyes-Gasga, T. Krekels, G. van Tendeloo, J. van Landuyt, S. Amelinckx, W.H.M. Bruggink and H. Verweij,
   Physica C \textbf{159}, 831 (1989).

\bibitem{Wer88}
   D.J. Werder, C.H. Chen, R.J. Cava and B. Batlogg,
   Phys. Rev. B \textbf{38}, 5130 (1988).

\bibitem{Bey89}
   R. Beyers, B.T. Ahn, G. Gorman, V.Y. Lee, S.S.P. Parkin, M.L. Ramirez, K.P. Roche, J.E. Vazquez,
   T.M. G\"ur and R.A. Huggins, Nature \textbf{340}, 619 (1989).

\bibitem{Ala88}
    M.A. Alario-Franco, C. Chaillout, J.J. Capponi, J. Chenavas and M. Marezio,
  Physica C \textbf{156},  455 (1988).

\bibitem{Son91}
   R. Sonntag, D. Hohlwein, T. Br\"uckel and G. Collin, Phys. Rev. Lett. \textbf{66}, 1497 (1991).

\bibitem{Zei92}
   Th. Zeiske, D. Hohlwein, R. Sonntag, F. Kubanek and Th. Wolf, Physica C \textbf{194}, 1 (1992).

\bibitem{Fle88}
   R.M. Flemming, L.F. Schneemeyer, P.K. Gallagher, B. Batlogg, L.W. Rupp and J.V. Waszczak,
   Phys. Rev. B \textbf{37}, 7920 (1988).

\bibitem{Pla94}
   V. Plakhty, B. Kviatkovsky, A. Stratilatov, Yu. Chernenkov, P. Burlet,
     J.Y. Henry, C. Marin, E. Ressouche, J. Schweizer, F. Yakou, E. Elkaim
     and J.P. Lauriat, Physica C \textbf{235-240}, 867 (1994).

\bibitem{Str98}
    E. Straube, D. Hohlwein and F. Kubanek, Physica C \textbf{295}, 1 (1998).

\bibitem{Pou96}
    H.F. Poulsen, M. von Zimmermann, J.R. Schneider, N.H. Andersen, P. Schleger, J. Madsen,
    R. Hadfield, H. Casalta, Ruixing Liang, P. Dosanjh and W.N. Hardy,
    Phys. Rev. B \textbf{53}, 15335 (1996).

\bibitem{Gry94}
   J. Grybos, D. Hohlwein, Th. Zeiske, R. Sonntag, F. Kubanek,
     K. Eichhorn and Th. Wolf, Physica C \textbf{220}, 138 (1994).

\bibitem{Bur92}
   P. Burlet, V.P. Plakthy, C. Marin and J.Y. Henry, Phys. Lett. A \textbf{167}, 401 (1992).

\bibitem{Zei91}
    Th. Zeiske, R. Sonntag, D. Hohlwein, N.H. Andersen and Th. Wolf,
    Nature \textbf{353}, 542 (1991).

\bibitem{Pla95}
    V. Plakhty, P. Burlet and J.Y. Henry, Physics Lett. A \textbf{198}, 256 (1995)

\bibitem{Sch94}
   W. Schweiss, W. Reichardt, M. Braden, G. Collin, G. Heger, H. Claus
     and A. Erb, Phys. Rev. B \textbf{49}, 1387 (1994).

\bibitem{Pla92}
    V. Plakhty, A. Startilov, Yu. Chernenkov, V. Fedorov, S.K. Sinha, Chun L. Loong, G.
    Gaulin, M. Vlasov and S. Moshkin,
    Sol. State Comm. \textbf{84}, 639 (1992).

\bibitem{Sch95a}
    P. Schleger, H. Casalta, R. Hadfield, H.F. Poulsen, M. von Zimmermann, N.H. Andersen,
    J.R. Schneider, R. Liang, P. Dosanjh and W.N. Hardy, Physica C \textbf{241}, 103 (1995).
    \label{ref:Sch95a}

\bibitem{Zei93}
    Th. Zeiske, D. Hohlwein, R. Sonntag, J. Grybos, K. Eichhorn and Th. Wolf,
    Physica C \textbf{207}, 333 (1993).

\bibitem{Ho94}
    D. Hohlwein in {\it Materials and Crystallographic Aspects of HTc-Superconductivity},
    edited by E. Kaldis (Kluwer Academic Publishers, The netherlands, 1994), p. 65.

\bibitem{Stra98}
    E. Straube, D. Hohlwein and F. Kubanek, Physica C \textbf{295}, 1 (1998).

\bibitem{Sch93}
   W. Scharz, O. Blaschko, G. Collin and F. Marucco, Phys. Rev. B  \textbf{48}, 6513 (1993).

\bibitem{Had94}
   R.A. Hadfield, P. Schleger, H. Casalta, N.H. Andersen,
   H.F. Poulsen, M. von Zimmermann, J.R. Schneider,
   M.T. Hutchings, D.A. Keen, Ruixing Liang, P. Dosanjh and W.N. Hardy,
   Physica C \textbf{235-240}, 1267 (1994).

\bibitem{Frel97}
    T. Frello, N.H. Andersen, J. Madsen, M. K\"all, M. von Zimmermann, O. Schmidt, H.F.
    Poulsen, J.R. Schneider and Th. Wolf, Physica C \textbf{282-287}, 1089 (1997).

\bibitem{Mon98}
    D. M{\o}nster, P.-A. Lindg{\aa}rd and N.H. Andersen, unpublished.

\bibitem{ref:Kal97}
    E. Kaldis, J. R\"{o}hler, E. Liarokapis, N. Poulakis, K. Conder and P.W. Loeffen,
   Phys. Rev. Lett. \textbf{79}, 4894 (1997).
   \label{ref:Kal97}

\bibitem{Liang92}
    R. Liang, P. Dosanjh, D.A. Bonn, D.J. Baar, J.F. Carolan and W.N. Hardy,
     Physica C \textbf{195}, 51 (1992).

\bibitem{Wolf89}
     Th. Wolf, W. Goldacker, B. Obst, G. Roth and R. Fl\"{u}kiger,
     J. Crys. Growth \textbf{96}, 1010 (1989).
     \label{ref:Wolf89}

\bibitem{Cas96}
    H. Casalta, P. Schleger, P. Harris, B. Lebech, N.H. Andersen, R. Liang,
    P. Dosanjh and W.N. Hardy, Physica C \textbf{258}, 321 (1996).

\bibitem{Bou98}
    R. Bouchard, D. Hupfeld, T. Lippmann, J. Neuefeind, H.-B. Neumann, H.F. Poulsen, U.
    R\"utt, T. Schmidt, J.R. Schneider, J. S\"ussenbach and M. von Zimmermann, Synchrotron
    Radiation News \textbf{5}, 90 (1998).

\bibitem{Neu96}
   H.-B. Neumann, J.R. Schneider, J. S\"ussenbach, S.R. Stock and Z.U. Rek,
   Nuclear Instrum. Methods A \textbf{372},  551 (1996).

\bibitem{Fii95}
    T. Fiig, N.H. Andersen, J. Berlin and P.A. Lindg{\aa}rd,
    Phys. Rev. B  \textbf{51}, 1226 (1995).

\bibitem{Bra94}
   A.J. Bray, Advances in Physics \textbf{43}, 357 (1994).

\bibitem{Cha87}
    C. Chaillout, M.A. Alario-Franco, J.J. Capponi, J. Chenavas, J.L. Hodeau and M.
    Marezio, Phys. Rev. B \textbf{36}, 7118 (1987).

\bibitem{Zei92a}
   Th. Zeiske, D. Hohlwein, R. Sonntag, F. Kubanek and G. Collin,
   Z. Phys. B  \textbf{86}, 11 (1992).

\bibitem{Kal98}
    M. K\"all, M. von Zimmermann, N.H. Andersen, J. Madsen, T. Frello, O. Schmidt,
    H.F. Poulsen, J.R. Schneider and Th. Wolf, unpublished.
    \label{ref:Kal98}

\bibitem{Brecht95}
    E. Brecht, W.W. Schmahl, H. Fuess, H. Casalta, P. Schleger, B. Lebech,
    N.H. Andersen and Th. Wolf, Phys. Rev. B \textbf{52}, 9601 (1995).

\bibitem{Frello99}
    T. Frello, M. Baziljevich, T.H. Johansen, N.H. Andersen and Th. Wolf,
     Phys. Rev. B \textbf{59}, R6639 (1999).

\bibitem{deF90}
   D. de Fontaine, G. Ceder and M. Asta, Nature \textbf{343}, 544 (1990).
    \label{ref:deF90}

\bibitem{Fiig96}
    T. Fiig, N.H. Andersen, P.-A. Lindg{\aa}rd, J. Berlin and O.G. Mouritsen, Phys Rev. B
    \textbf{54}, 556 (1996).

\bibitem{deF87}
   D. de Fontaine, L.T. Wille and S.C. Moss,
   Phys. Rev. B \textbf{36}, 5709  (1987).

\bibitem{Ste89}
    P.A. Sterne and L.T. Wille, Physica C \textbf{162-164}, 223 (1989).

\bibitem{Kha90}
    A.G. Khachaturyan and J.W. Morris, Phys. Rev. Lett. \textbf{64}, 1989 (1990).

\bibitem{Ber89}
    A. Bertinotti, J. Hammann, D. Luzet and A. Vinzent,
     Physica C \textbf{160}, 227 (1989).

\bibitem{Yak96}
   F. Yakhou, V. Plakhty, A. Stratilatov, P. Burlet, J.P. Lauriat, E. Elkaim, J.Y. Henry, M.
   Vlasov and S. Moshkin, Physica C  \textbf{261}, 315 (1996).

\bibitem{Kre90}
   T. Krekels, T.S. Shi, J. Reyes-Gasga, G. van Tendeloo, J. van Landuyt
   and S. Amelinckx, Physica C \textbf{167}, 677 (1990).

\bibitem{Kre95}
   T. Krekels, S. Kaesche and G. van Tendeloo, Physica C \textbf{248}, 317 (1995).

\bibitem{Wer91}
    D.J. Werder, C.H. Chen and G.P. Espinosa, Physica C \textbf{173}, 285 (1991).

\bibitem{Cava90}
    R.J. Cava, A.W. Hewat, E.A. Hewat, B. Batlogg, M. Marezio, K.M. Rabe, J.J. Krajewski,
    W.F. Peck Jr. and L.W. Rupp Jr., Physica C \textbf{165}, 419 (1990).
    \label{ref:Cava90}

\bibitem{Tol92}
    H. Tolentino, F. Baudelet, A. Fontaine, T. Gourieux, G. Krill, J.Y. Henry and J.
    Rossat-Mignod, Physica C \textbf{192}, 115 (1992).

\bibitem{Lue96}
    H. L{\"u}etgemeier, S. Schmenn, P. Meuffels, O. Storz, R. Sch{\"o}lhorn, Ch.
    Niedermeier, I. Heinmaa and Yu. Baikov, Physica C \textbf{267}, 191 (1996).

\bibitem{Pou91}
    H.F. Poulsen, N.H. Andersen, J.V. Andersen, H. Bohr and O.G. Mouritsen, Nature
    \textbf{349}, 594 (1991).

\bibitem{Pou91a}
    H.F. Poulsen, N.H. Andersen, J.V. Andersen, H. Bohr and O.G. Mouritsen, Phys. Rev.
    Lett. \textbf{66}, 465 (1991).

\bibitem{Zaa88}
    J. Zaanen, A.T. Paxton, O. Jepsen and O.K. Andersen, Phys Rev. Lett. \textbf{60},
    2685 (1988).

\bibitem{McC92}
    R. McCormack, D. de Fontaine and G. Ceder, Phys. Rev. B \textbf{45}, 12976 (1992).

\bibitem{Uim94}
    G. Uimin, Phys. Rev. B \textbf{50}, 9531 (1994).

\bibitem{Ali94}
    A.A. Aligia and J. Garc\'{e}s, Phys. Rev. B \textbf{49}, 524 (1994).

\bibitem{Hau97}
    H. Haugerud, G. Uimin and W. Selke, Physica C \textbf{275}, 93 (1997).

\bibitem{Ali90}
    A.A. Aligia, J. Garc\'{e}s and H. Bonadeo, Phys. Rev. B \textbf{42}, 10226 (1990).

\bibitem{Sem92}
    S. Semenovska and A.G. Khachaturian, Phys. Rev. B \textbf{46}, 6511 (1992).

\bibitem{Kha88}
    A.G. Khachaturian and J.W. Morris Jr., Phys. Rev. Lett. \textbf{61}, 215 (1988).

\bibitem{Lin97}
   P.A. Lindg{\aa}rd, N.H. Andersen, D. M{\o}nster and T. Fiig, In: Scientific Computing
   Report 1995-1997, J. Wasniewski (Ed.), (UNI-C, Copenhagen, 1998), 38 (1998).

\bibitem{Ali94a}
    A.A. Aligia, J. Garc\'{e}s and J.P. Abriata, Physica C \textbf{221}, 109 (1994).

\end{thebibliography}
\end{document}